\documentclass[12pt]{article}
\usepackage[cp1250]{inputenc}
\usepackage{amsmath,amssymb,amsthm}

\def\MB{\mathbf}
\def\MC{\mathcal}
\def\MR{\mathrm}
\def\MF{\mathfrak}
\def\BB{\mathbb}
\def\R{\BB{R}}

\def\wh{\widehat}
\def\ol{\overline}

\def\Tr{\operatorname{\MR{Tr}}\nolimits}
\def\tsum{\textstyle\sum\limits}
\def\tprod{\textstyle\prod\limits}

\def\slacs#1{\setlength{\arraycolsep}{#1}}

\slacs{.4ex}
\begin{document}

\begin{center}
\Large\textbf{Nested Bethe Ansatz for RTT--Algebra $\MC{A}_n$}
\end{center}

\centerline{\v{C}. Burd\'{\i}k\,$^{1)}$, O. Navr\'atil\,$^{2)}$}

\vskip5mm

\centerline{$^{1)}$\,Faculty of Nuclear Sciences and Physical Engineering, CTU,}
\centerline{Trojanova 13, Prague, Czech Republic}
\centerline{email: burdices@kmlinux.fjfi.cvut.cz}

\vskip3mm

\centerline{$^{2)}$\,Faculty of Transportation Sciences, CTU,}
\centerline{Na Florenci 25, Prague, Czech Republic}
\centerline{email: navraond@fd.cvut.cz}

\begin{abstract}
This paper continues our recent studies on the algebraic Bethe ansatz
for the RTT--algebras of $\MR{sp}(2n)$ and $\MR{o}(2n)$ types.
In these studies, we encountered the RTT--algebras which we called $\MC{A}_n$.
The next step in our construction of the Bethe vectors for the RTT-algebras
of type $\MR{sp}(2n)$ and $\MR{o}(2n)$ is to find the Bethe vectors for
the RTT--algebras $\MC{A}_n$. This paper deals with the construction of the
Bethe vectors of the RTT--algebra $\MC{A}_n$ using the Bethe vectors
of the RTT--algebra $\MC{A}_{n-1}$.
\end{abstract}

\section{Introduction}
 \label{uvod}

In studying the algebraic Bethe ansatz for the RTT--algebras of type
$\MR{sp}(2n)$ and $\MR{o}(2n)$ \cite{BN-SP, BN-SO}, we discovered
some the RTT--algebras which we called $\MC{A}_n$.
The main result of these works is the assertion that
for the construction of eigenvalues and eigenvectors of the transfer--matrix
of the RTT--algebras of type $\MR{sp}(2n)$ and $\MR{o}(2n)$ it is
enough to find eigenvalues and eigenvectors for the RTT--algebra $\MC{A}_n$.

In this work, we deal with the nested Bethe ansatz for
the RTT--algebra $\MC{A}_n$. We show how to construct
eigenvectors for the RTT--algebra $\MC{A}_n$ by using eigenvectors
of the RTT--algebra $\MC{A}_{n-1}$.

Note, that the RTT--algebra $\MC{A}_{n-1}$ is not the
RTT--subalgebra $\MC{A}_n$. However, $\MC{A}_n$ contains two the
RTT--subalgebras $\MC{A}^{(+)}_n$ and $\MC{A}^{(-)}_n$, which are of
type $\MR{gl}(n)$. The RTT--algebras $\MC{A}^{(\pm)}_{n-1}$ are
already the RTT--subalgebras of $\MC{A}^{(\pm)}_n$. As we will see
later, we can construct some eigenvectors for the RTT--algebras
$\MC{A}_n$ as Bethe vectors of the RTT--algebras $\MC{A}^{(\pm)}_n$,
i.e. as the Bethe vectors for the RTT--algebras of the type
$\MR{gl}(n)$. Our result for such eigenvectors is the same as for
the nested Bethe ansatz for the RTT--algebras of $\MR{gl}(n)$, which
can be found in \cite{KR83}. In this sense, our construction is a
certain generalization of the nested Bethe ansatz for the
RTT--algebras of type $\MR{gl}(n)$.

The proofs of many claims are only a suitable, but long adjustment
of the Yang--Baxter and the RTT--equations. We have included them in Appendix
for better clarity of the main text.

\section{ The RTT--algebra $\MC{A}_n$}
 \label{A-n}

We denote $\MB{E}^i_k$ and $\MB{E}^{-i}_{-k}$, where $i,k=1,\,\ldots,n$,
the matrices
$\bigl(\MB{E}^i_k\bigr)^r_s=\bigl(\MB{E}^{-i}_{-k}\bigr)^{-r}_{-s}=
\delta^r_k\delta^i_s$. Then the relations
$\MB{E}^i_k\MB{E}^r_s=\delta^i_s\MB{E}^r_k$,
${\tsum_{i=1}^n}\MB{E}^i_i=\MB{I}_+$ and
${\tsum_{i=1}^n}\MB{E}^{-i}_{-i}=\MB{I}_-$
apply.

The RTT--algebra $\MC{A}_n$ is an associative algebra with a unit that is
generated by the elements $T^i_k(x)$ and $T^{-i}_{-k}(x)$, where $i,k=1,\ldots,n$.
If we introduce the monodromy matrix
$\MB{T}(x)=\MB{T}^{(+)}(x)+\MB{T}^{(-)}(x)$, where
$$
\MB{T}^{(+)}(x)={\tsum_{i,k=1}^n}\MB{E}^k_i\otimes T^i_k(x)\,,\qquad
\MB{T}^{(-)}(x)={\tsum_{i,k=1}^n}\MB{E}^{-k}_{-i}\otimes
T^{-i}_{-k}(x)\,,
$$
the commutation relations between generators are defined by
the RTT--equation
\begin{equation}
 \label{RTT-1}
\MB{R}_{1,2}(x,y)\MB{T}_1(x)\MB{T}_2(y)=
\MB{T}_2(y)\MB{T}_1(x)\MB{R}_{1,2}(x,y),
\end{equation}
where R--matrix is
$\MB{R}(x,y)=\MB{R}^{(+,+)}(x,y)+\MB{R}^{(+,-)}(x,y)+
\MB{R}^{(-,+)}(x,y)+\MB{R}^{(-,-)}(x,y)$,
$$
\begin{array}{l}
\MB{R}^{(+,+)}(x,y)= \dfrac1{f(x,y)}\,\Bigl(\MB{I}_+\otimes\MB{I}_++
g(x,y){\tsum_{i,k=1}^n}\MB{E}^i_k\otimes\MB{E}^k_i\Bigr),\\[6pt]
\MB{R}^{(+,-)}(x,y)= \MB{I}_+\otimes\MB{I}_--
k(x,y){\tsum_{i,k=1}^n}\MB{E}^i_k\otimes\MB{E}^{-i}_{-k}\,,\\[6pt]
\MB{R}^{(-,+)}(x,y)= \MB{I}_-\otimes\MB{I}_+-
h(x,y){\tsum_{i,k=1}^n}\MB{E}^{-i}_{-k}\otimes\MB{E}^i_k\,,\\[6pt]
\MB{R}^{(-,-)}(x,y)= \dfrac1{f(x,y)}\,\Bigr(\MB{I}_-\otimes\MB{I}_-+
g(x,y){\tsum_{i,k=1}^n}\MB{E}^{-i}_{-k}\otimes\MB{E}^{-k}_{-i}\Bigr),\\[6pt]
g(x,y)=\dfrac1{x-y}\,,\hskip22mm f(x,y)=\dfrac{x-y+1}{x-y}\,,\\[6pt]
h(x,y)=\dfrac1{x-y+n-\eta}\,,\qquad k(x,y)=\dfrac1{x-y+\eta}
\end{array}
$$
and $\eta$ is any number. For $\eta=-1$ we obtain the RTT--algebra
connected with the RTT--algebra of $\MR{sp}(2n)$ type and for $\eta=1$
the RTT--algebra connected with the RTT-algebra of $\MR{o}(2n)$ type.

By direct calculation, it can be verified that this R--matrix
satisfies the Yang--Baxter equation
\begin{equation}
 \label{YB-1}
\MB{R}_{1,2}(x,y)\MB{R}_{1,3}(x,z)\MB{R}_{2,3}(y,z)=
\MB{R}_{2,3}(y,z)\MB{R}_{1,3}(x,z)\MB{R}_{1,2}(x,y)
\end{equation}
and has the inverse R--matrix
$$
\bigl(\MB{R}(x,y)\bigr)^{-1}= \bigl(\MB{R}^{(+,+)}(x,y)\bigr)^{-1}+
\bigl(\MB{R}^{(+,-)}(x,y)\bigr)^{-1}+
\bigl(\MB{R}^{(-,+)}(x,y)\bigr)^{-1}+
\bigl(\MB{R}^{(-,-)}(x,y)\bigr)^{-1}
$$
where
$$
\begin{array}{l}
\bigl(\MB{R}^{(+,+)}(x,y)\bigr)^{-1}=
\dfrac1{f(y,x)}\,\Bigl(\MB{I}_+\otimes\MB{I}_++
g(y,x){\tsum_{i,k=1}^n}\MB{E}^i_k\otimes\MB{E}^k_i\Bigr),\\[6pt]
\bigl(\MB{R}^{(+,-)}(x,y)\bigr)^{-1}= \MB{I}_+\otimes\MB{I}_--
h(y,x){\tsum_{i,k=1}^n}\MB{E}^i_k\otimes\MB{E}^{-i}_{-k}\,,\\[6pt]
\bigl(\MB{R}^{(-,+)}(x,y)\bigr)^{-1}= \MB{I}_-\otimes\MB{I}_+-
k(y,x){\tsum_{i,k=1}^n}\MB{E}^{-i}_{-k}\otimes\MB{E}^i_k\,,\\[6pt]
\bigl(\MB{R}^{(-,-)}(x,y)\bigr)^{-1}=
\dfrac1{f(y,x)}\,\Bigl(\MB{I}_-\otimes\MB{I}_-+
g(y,x){\tsum_{i,k=1}^n}\MB{E}^{-i}_{-k}\otimes\MB{E}^{-k}_{-i}\Bigr)\,.
\end{array}
$$
Therefore, it defines the RTT--algebra that we denote by $\MC{A}_n$.

The explicit form of commutation relations between generators
of the RTT--algebra $\MC{A}_n$ is given in the Appendix.

It is easily seen that the RTT--equation (\ref{RTT-1}) can be written as
\begin{equation}
 \label{RTT-pm}
\MB{R}^{(\epsilon_1,\epsilon_2)}_{1,2}(x,y)
\MB{T}^{(\epsilon_1)}_1(x)\MB{T}^{(\epsilon_2)}_2(y)=
\MB{T}^{(\epsilon_2)}_2(y)\MB{T}^{(\epsilon_1)}_1(x)
\MB{R}^{(\epsilon_1,\epsilon_2)}_{1,2}(x,y)\,,
\end{equation}
where $\epsilon_1,\,\epsilon_2=\pm$. From this form of the RTT--equation
it is clear that in the RTT--algebra $\MC{A}_n$ there are two RTT-subalgebras
$\MC{A}^{(+)}_n$ and $\MC{A}^{(-)}_n$, which are generated by the elements
$T^i_k(x)$ and $T^{-i}_{-k}(x)$, where $i,\,k=1,\,\ldots,\,n$.

Using the RTT--equation (\ref{RTT-pm}), it is possible to show
that in the RTT--algebra $\MC{A}_n$ the operators
$$
H^{(+)}(x)=\Tr\MB{T}^{(+)}(x)={\tsum_{i=1}^n}T^i_i(x)\,,\qquad
H^{(-)}(x)=\Tr\MB{T}^{(-)}(x)={\tsum_{i=1}^n}T^{-i}_{-i}(x)\,
$$
mutually commute.

We deal with the representations of the RTT--algebra $\MC{A}_n$
on the vector space $\MC{W}=\MC{A}_n\omega$, where $\omega$
is a vacuum vector for which the relations
$$
\begin{array}{lcll}
T^i_k(x)\omega=0 &\quad\MR{for}\quad& 1\leq i<k\leq n\,,\hskip10mm&
T^i_i(x)\omega=\lambda_i(x)\,,\omega\\[6pt]
T^{-k}_{-i}(x)\omega=0 &\quad\MR{for}\quad& 1\leq i<k\leq
n\,,\hskip10mm& T^{-i}_{-i}(x)\omega=\lambda_{-i}(x)\omega
\end{array}
$$
hold. Our goal is to find in the vector space $\MC{W}$
common eigenvectors of the operators $H^{(\pm)}(x)$.

\bigskip

In the RTT--algebra $\MC{A}_n$ there are two RTT--subalgebras
$\tilde{\MC{A}}^{(+)}=\MC{A}^{(+)}_{n-1}$ and
$\tilde{\MC{A}}^{(-)}=\MC{A}^{(-)}_{n-1}$ of $\MR{gl}(n-1)$ type,
which are generated by the elements $T^i_k(x)$ and
$T^{-i}_{-k}(x)$, where $i,\,k=1,\,\ldots,n-1$.

First, we will consider the subspace $\tilde{\MC{W}}$ generated
by the elements $\tilde{\MC{A}}^{(+)}\tilde{\MC{A}}^{(-)}\omega$.

\bigskip

\underbar{\textbf{Proposition 1.}}
The relations
\begin{equation}
 \label{L1}
T^i_n(x)w=T^{-n}_{-i}(x)w=0\,,\qquad
T^n_n(x)w=\lambda_n(x)w\,,\qquad
T^{-n}_{-n}(x)w=\lambda_{-n}(x)w
\end{equation}
hold for any $w\in\tilde{\MC{W}}$ and $i=1,\,2,\,\ldots,\,n-1$.

\medskip

\textsc{Proof.} First, we consider the space
$\tilde{\MC{W}}^{(-)}=\tilde{\MC{A}}^{(-)}\omega\subset\tilde{\MC{W}}$.
To prove relation (\ref{L1}) for $w=w^{(-)}\in\MC{W}^{(-)}$, it is
sufficient to show that if (\ref{L1}) is valid for $w^{(-)}$, it
also applies to $T^{-r}_{-s}(y)w^{(-)}$, where $r,s=1,\ldots,n-1$.
From the commutation relations we get for $i=1,\,\ldots,\,n$ and
$r,\,s=1,\,\ldots,\,n-1$
$$
T^{-n}_{-i}(x)T^{-r}_{-s}(y)= T^{-r}_{-s}(y)T^{-n}_{-i}(x)+
g(x,y)T^{-r}_{-i}(y)T^{-n}_{-s}(x)-g(x,y)T^{-r}_{-i}(x)T^{-n}_{-s}(y)
$$
It follows that for any $w^{(-)}\in\tilde{\MC{W}}^{(-)}$ we have
$$
T^{-n}_{-i}(x)w^{(-)}=0\quad\MR{for}\quad i=1,\,\ldots,\,n-1\,,
\hskip10mm T^{-n}_{-n}(x)w^{(-)}=\lambda_{-n}(x)w^{(-)}\,.
$$
For any $r,\,s=1,\,\ldots,\,n-1$, the commutation relations give
$$
T^n_n(x)T^{-r}_{-s}(y)=T^{-r}_{-s}(y)T^n_n(x)\,,
$$
which proves that $T^n_n(x)w^{(-)}=\lambda_n(x)w^{(-)}$ for any
$w^{(-)}\in\tilde{\MC{W}}^{(-)}$.

For any $i,\,r,\,s=1,\,\ldots,\,n-1$ the relations
$$
T^i_n(x)T^{-r}_{-s}(y)= T^{-r}_{-s}(y)T^i_n(x)-
\delta^{i,r}h(y,x){\tsum_{p=1}^{n-1}}T^{-p}_{-s}(y)T^p_n(x)-
\delta^{i,r}h(y,x)T^{-n}_{-s}(y)T^n_n(x).
$$
hold. Since for every $w^{(-)}\in\tilde{\MC{W}}^{(-)}$
$$
T^{-n}_{-s}(y)T^n_n(x)w^{(-)}=\lambda_n(x)T^{-n}_{-s}(y)w^{(-)}=0\,,
$$
we see that for every $w^{(-)}\in\tilde{\MC{W}}^{(-)}$ and
$i=1,\,\ldots,\,n-1$ we have $T^i_n(x)w^{(-)}=0$.

Since $\tilde{\MC{W}}=\MC{A}^{(+)}\tilde{\MC{W}}^{(-)}$,
it is sufficient to show that if (\ref{L1}) holds for $w$,
it also holds for $T^r_s(y)w$, where $r,s=1,\ldots,n-1$.

For $i=1,\,\ldots,\,n$ a $r,\,s=1,\,\ldots,\,n-1$ we have
the commutation relation
$$
T^i_n(x)T^r_s(y)=
T^r_s(y)T^i_n(x)+g(y,x)T^i_s(y)T^r_n(x)-g(y,x)T^i_s(x)T^r_n(y)\,,
$$
from which we can easily see that for any $w\in\tilde{\MC{W}}$
$$
T^i_n(x)w=0\quad\MR{for}\quad i=1,\,\ldots,\,n-1\,,\qquad
T^n_n(x)w=\lambda_n(x)w
$$
holds.

The relation $T^{-n}_{-n}(x)w=\lambda_{-n}(x)w$ results from the fact
that for every $r,\,s=1,\,\ldots,\,n-1$ we have
$$
T^{-n}_{-n}(x)T^r_s(y)=T^r_s(y)T^{-n}_{-n}(x)\,.
$$
For $i,\,r,\,s=1,\,\ldots,\,n-1$ we use
$$
T^{-n}_{-i}(x)T^r_s(y)=T^r_s(y)T^{-n}_{-i}(x)-
\delta_{i,s}h(x,y){\tsum_{p=1}^{n-1}}T^r_p(y)T^{-n}_{-p}(x)-
\delta_{i,s}h(x,y)T^r_n(y)T^{-n}_{-n}(x)\,,
$$
which implies that $T^{-n}_{-i}(x)w=0$ for $i=1,\,\ldots,n-1$ and
for any $w\in\tilde{\MC{W}}$. \qed

\bigskip

\underbar{\textbf{Proposition 2.}} The space $\tilde{\MC{W}}$ is
invariant with respect to $\tilde{\MC{A}}^{(+)}$ and
$\tilde{\MC{A}}^{(-)}$.

\smallskip

\textsc{Proof:} Obviously, the space $\tilde{\MC{W}}$ is
invariant for the $\tilde{\MC{A}}^{(+)}$ action.

To show that the space $\tilde{\MC{W}}$ is invariant to the
action of the algebra $\tilde{\MC{A}}^{(-)}$, we will use for
$i,\,k,\,r,\,s=1,\,\ldots,\,n-1$ the commutation relations
$$
\begin{array}{l}
T^{-i}_{-k}(x)T^r_s(y)-
\delta_{k,s}k(y,x){\tsum_{p=1}^{n-1}}T^{-i}_{-p}(x)T^r_p(y)-
\delta_{k,s}k(y,x)T^{-i}_{-n}(x)T^r_n(y)=\\[6pt]
\hskip20mm= T^r_s(y)T^{-i}_{-k}(x)-
\delta^{i,r}k(y,x){\tsum_{p=1}^{n-1}}T^p_s(y)T^{-p}_{-k}(x)-
\delta^{i,r}k(y,x)T^n_s(y)T^{-n}_{-k}(x)\,.
\end{array}
$$
From Proposition 1 it follows that if we restrict these relations to subspace
$\tilde{\MC{W}}$, we get
$$
T^{-i}_{-k}(x)T^r_s(y)-
\delta_{k,s}k(y,x){\tsum_{p=1}^{n-1}}T^{-i}_{-p}(x)T^r_p(y)=
T^r_s(y)T^{-i}_{-k}(x)-
\delta^{i,r}k(y,x){\tsum_{p=1}^{n-1}}T^p_s(y)T^{-p}_{-k}(x)\,.
$$
If we multiply these equations by
$\bigl(\delta^k_a\delta^s_b-\delta^{k,s}\delta_{a,b}\tilde{h}(x,y)\bigr)$,
where
$$
\tilde{h}(x,y)=\frac1{x-y+(n-1)-\eta}\,,
$$
and sum them over $k$, $s$ from 1 to $n-1$ and rename the indices, we find
that the relations
$$
\begin{array}{l}
T^{-i}_{-k}(x)T^r_s(y)= T^r_s(y)T^{-i}_{-k}(x)-
\delta^{i,r}k(y,x){\tsum_{p=1}^{n-1}}T^p_s(y)T^{-p}_{-k}(x)-\\[6pt]
\hskip25mm-
\delta_{k,s}\tilde{h}(x,y){\tsum_{p=1}^{n-1}}T^r_p(y)T^{-i}_{-p}(x)+
\delta^{i,r}\delta_{k,s}\tilde{h}(x,y)k(y,x){\tsum_{p,q=1}^{n-1}}T^p_q(y)T^{-p}_{-q}(x)
\end{array}
$$
are true on the space $\tilde{\MC{W}}$.

The invariance of the space $\tilde{\MC{W}}$ with respect to the action
$\tilde{\MC{A}}^{(-)}$ can be proven by induction according to numbers
of the factors $T^i_k(y)$ in the vectors $w\in\tilde{\MC{W}}$. \qed

\bigskip

\underbar{\textbf{Proposition 3.}} If we define
$$
\tilde{\MB{T}}^{(+)}(x)={\tsum_{i,k=1}^{n-1}}\MB{E}^k_i\otimes T^i_k(x)\,,
\qquad
\tilde{\MB{T}}^{(-)}(x)={\tsum_{i,k=1}^{n-1}}\MB{E}^{-k}_{-i}\otimes T^{-i}_{-k}(x)
$$
the commutation relations for $T^i_k(x)$ and $T^{-i}_{-k}(x)$, where
$i,\,k=1,\,\ldots,\,n-1$, reduced to the space $\tilde{\MC{W}}$
can been written in the form of the RTT--equation
$$
\tilde{\MB{R}}^{(\epsilon_1,\epsilon_2)}_{1,2}(x,y)
\tilde{\MB{T}}^{(\epsilon_1)}_1(x)
\tilde{\MB{T}}^{(\epsilon_2)}_2(y)=
\tilde{\MB{T}}^{(\epsilon_2)}_2(y)
\tilde{\MB{T}}^{(\epsilon_1)}_1(x)
\tilde{\MB{R}}^{(\epsilon_1,\epsilon_2)}_{1,2}(x,y)
$$
where $\epsilon_1,\,\epsilon_2=\pm$ and
$$
\begin{array}{l}
\tilde{\MB{R}}^{(+,+)}_{1,2}(x,y)=
\dfrac1{f(x,y)}\,\Bigl(\tilde{\MB{I}}_+\otimes\tilde{\MB{I}}_++
g(x,y){\tsum_{i,k=1}^{n-1}}\MB{E}^i_k\otimes\MB{E}^k_i\Bigr),\\[6pt]
\tilde{\MB{R}}^{(-,-)}_{1,2}(x,y)=
\dfrac1{f(x,y)}\,\Bigr(\tilde{\MB{I}}_-\otimes\tilde{\MB{I}}_-+
g(x,y){\tsum_{i,k=1}^{n-1}}\MB{E}^{-i}_{-k}\otimes\MB{E}^{-k}_{-i}\Bigr),\\[6pt]
\tilde{\MB{R}}^{(+,-)}_{1,2}(x,y)=
\tilde{\MB{I}}_+\otimes\tilde{\MB{I}}_--
k(x,y){\tsum_{i,k=1}^{n-1}}\MB{E}^i_k\otimes\MB{E}^{-i}_{-k},\\[6pt]
\tilde{\MB{R}}^{(-,+)}_{1,2}(x,y)=
\tilde{\MB{I}}_-\otimes\tilde{\MB{I}}_+-
\tilde{h}(x,y){\tsum_{i,k=1}^{n-1}}\MB{E}^{-i}_{-k}\otimes\MB{E}^i_k\\[6pt]
\tilde{\MB{I}}_+={\tsum_{k=1}^{n-1}}\MB{E}^k_k\,,\qquad
\tilde{\MB{I}}_-={\tsum_{k=1}^{n-1}}\MB{E}^{-k}_{-k}\,,\qquad
\tilde{h}(x,y)=\dfrac1{x-y+n-1-\eta}\,.
\end{array}
$$

\textsc{Proof:} If we consider only the indices
$i,\,k,\,r,\,s=1,\,\ldots,\,n-1$ in the commutation relations
$$
\begin{array}{l}
T^i_k(x)T^r_s(y)+g(x,y)T^r_k(x)T^i_s(y)=
T^r_s(y)T^i_k(x)+g(x,y)T^r_k(y)T^i_s(x)\\[6pt]
T^{-i}_{-k}(x)T^{-r}_{-s}(y)+g(x,y)T^{-r}_{-k}(x)T^{-i}_{-s}(y)=
T^{-r}_{-s}(y)T^{-i}_{-k}(x)+g(x,y)T^{-r}_{-k}(y)T^{-i}_{-s}(x)
\end{array}
$$
we can write them in the matrix form
$$
\begin{array}{l}
\tilde{\MB{R}}^{(+,+)}_{1,2}(x,y) \tilde{\MB{T}}^{(+)}_1(x)
\tilde{\MB{T}}^{(+)}_2(y)= \tilde{\MB{T}}^{(+)}_2(y)
\tilde{\MB{T}}^{(+)}_1(x)
\tilde{\MB{R}}^{(+,+)}_{1,2}(x,y)\\[4pt]
\tilde{\MB{R}}^{(-,-)}_{1,2}(x,y) \tilde{\MB{T}}^{(-)}_1(x)
\tilde{\MB{T}}^{(-)}_2(y)= \tilde{\MB{T}}^{(-)}_2(y)
\tilde{\MB{T}}^{(-)}_1(x) \tilde{\MB{R}}^{(-,-)}_{1,2}(x,y)\,.
\end{array}
$$

For $i,\,k,\,r,\,s=1,\,\ldots,\,n-1$ we have in the RTT--algebra $\MC{A}_n$
the commutation relations
$$
\begin{array}{l}
T^i_k(x)T^{-r}_{-s}(y)-\delta^{i,r}k(x,y){\tsum_{p=1}^n}T^p_k(x)T^{-p}_{-s}(y)=
T^{-r}_{-s}(y)T^i_k(x)-\delta_{k,s}k(x,y){\tsum_{p=1}^n}T^{-r}_{-p}(y)T^i_p(x)\\[6pt]
T^{-i}_{-k}(x)T^r_s(y)-\delta_{k,s}k(y,x){\tsum_{p=1}^n}T^{-i}_{-p}(x)T^r_p(y)=
T^r_s(y)T^{-i}_{-k}(x)-\delta^{i,r}k(y,x){\tsum_{p=1}^n}T^p_s(y)T^{-p}_{-k}(x)\,.
\end{array}
$$
If we restrict them on the space $\tilde{\MC{W}}$,
we obtain according to Proposition 1
$$
\begin{array}{l}
T^i_k(x)T^{-r}_{-s}(y)-\delta^{i,r}k(x,y){\tsum_{p=1}^{n-1}}T^p_k(x)T^{-p}_{-s}(y)=
T^{-r}_{-s}(y)T^i_k(x)-\delta_{k,s}k(x,y){\tsum_{p=1}^{n-1}}T^{-r}_{-p}(y)T^i_p(x)\\[6pt]
T^{-i}_{-k}(x)T^r_s(y)-\delta_{k,s}k(y,x){\tsum_{p=1}^{n-1}}T^{-i}_{-p}(x)T^r_p(y)=
T^r_s(y)T^{-i}_{-k}(x)-\delta^{i,r}k(y,x){\tsum_{p=1}^{n-1}}T^p_s(y)T^{-p}_{-k}(x)\,.
\end{array}
$$
The first of these commutation relations is
$$
\tilde{\MB{R}}^{(+,-)}_{1,2}(x,y) \tilde{\MB{T}}^{(+)}_1(x)
\tilde{\MB{T}}^{(-)}_2(y)= \tilde{\MB{T}}^{(-)}_2(y)
\tilde{\MB{T}}^{(+)}_1(x) \tilde{\MB{R}}^{(+,-)}_{1,2}(x,y).
$$
The second equality can be written using matrices in the form
$$
\begin{array}{l}
\tilde{\MB{T}}^{(-)}_1(x)\tilde{\MB{T}}^{(+)}_2(y)
\Bigl(\tilde{\MB{I}}_-\otimes\tilde{\MB{I}}_+-
k(y,x){\tsum_{i,k=1}^{n-1}}\MB{E}^{-i}_{-k}\otimes\MB{E}^i_k\Bigr)=\\[6pt]
\hskip20mm= \Bigl(\tilde{\MB{I}}_-\otimes\tilde{\MB{I}}_+-
k(y,x){\tsum_{i,k=1}^{n-1}}\MB{E}^{-i}_{-k}\otimes\MB{E}^i_k\Bigr)
\tilde{\MB{T}}^{(+)}_2(y)\tilde{\MB{T}}^{(-)}_1(x).
\end{array}
$$
And since
$$
\begin{array}{l}
\tilde{\MB{R}}^{(-,+)}_{1,2}(x,y)
\Bigl(\tilde{\MB{I}}_-\otimes\tilde{\MB{I}}_+-
k(y,x){\tsum_{i,k=1}^{n-1}}\MB{E}^{-i}_{-k}\otimes\MB{E}^i_k\Bigr)=\\[4pt]
\hskip20mm= \Bigl(\tilde{\MB{I}}_-\otimes\tilde{\MB{I}}_+-
k(y,x){\tsum_{i,k=1}^{n-1}}\MB{E}^{-i}_{-k}\otimes\MB{E}^i_k\Bigr)
\tilde{\MB{R}}^{(-,+)}_{1,2}(x,y)=\tilde{\MB{I}}_-\otimes\tilde{\MB{I}}_+
\end{array}
$$
this relation is equivalent to the RTT--equation
$$
\tilde{\MB{R}}^{(-,+)}_{1,2}(x,y)
\tilde{\MB{T}}^{(-)}_1(x)\tilde{\MB{T}}^{(+)}_2(y)=
\tilde{\MB{T}}^{(-)}_2(y)\tilde{\MB{T}}^{(+)}_1(x)
\tilde{\MB{R}}^{(-,+)}_{1,2}(x,y)\,.
$$
\qed

\bigskip

The following theorem immediately follows from Proposition 3.

\smallskip

\underbar{\textbf{Theorem 1.}} The action of the operators $T^i_k(x)$ and
$T^{-i}_{-k}(x)$, where $i,\,k=1,\,\ldots,\,n-1$, in the space
$\tilde{\MC{W}}$ forms the RTT--algebra $\MC{A}_{n-1}$.

\section{General form of common eigenvectors of
$H^{(+)}(x)$ and $H^{(-)}(x)$}.
 \label{obecny}

Let $\vec{v}=(v_1,v_2,\ldots,v_P)$ and $\vec{w}=(w_1,w_2,\ldots,w_Q)$
be ordered sets of mutually different numbers.
 We will search for a general shape of the common eigenvectors $H^{(+)}(x)$ and
 $H^{(-)}(x)$ in the form
$$
\MF{B}(\vec{v},\vec{w})={\tsum_{k_1,\ldots,k_P=1}^{n-1}}\,{\tsum_{r_1,\ldots,r_Q=1}^{n-1}}
T^n_{k_1}(v_1)\ldots T^n_{k_P}(v_P) T^{-r_1}_{-n}(w_1)\ldots
T^{-r_Q}_{-n}(w_Q)\Phi^{k_1,\ldots,k_P}_{-r_1,\ldots,-r_Q}\,,
$$
where $\Phi^{k_1,\ldots,k_P}_{-r_1,\ldots,-r_Q}\in\tilde{\MC{W}}$.

We will consider $(n-1)$--dimensional spaces $\MC{V}_+$ and
$\MC{V}_-$ with the base $\MB{e}_k$ and $\MB{e}_{-r}$ and denote
$\MB{f}^k$ and $\MB{f}^{-r}$ their dual base in dual spaces
$\MC{V}^*_+$ and $\MC{V}^*_-$.

Let us define
$$
\begin{array}{l}
\MB{b}^{(+)}(v)={\tsum_{k=1}^{n-1}}\MB{f}^k\otimes T^n_k(v)\in
\MC{V}^*_+\otimes\MC{A}_n\\[6pt]
\MB{b}^{(-)}(w)={\tsum_{r=1}^{n-1}}\MB{e}_{-r}\otimes
T^{-r}_{-n}(w)\in \MC{V}_-\otimes\MC{A}_n
\end{array}
$$
and denote
$$
\begin{array}{l}
\MB{b}^{(+)}_{1^*,\ldots,P^*}(\vec{v})=
\MB{b}^{(+)}_{1^*}(v_1)\MB{b}^{(+)}_{2^*}(v_2)\ldots\MB{b}^{(+)}_{P^*}(v_P)\in
\MC{V}^*_1\otimes\MC{V}^*_2\otimes\ldots\otimes\MC{V}^*_P\otimes\MC{A}_n\\[4pt]
\MB{b}^{(-)}_{1,\ldots,Q}(\vec{w})=
\MB{b}^{(-)}_1(w_1)\MB{b}^{(-)}_2(w_2)\ldots\MB{b}^{(-)}_Q(w_Q)\in
\MC{V}_{-1}\otimes\MC{V}_{-2}\otimes\ldots\otimes\MC{V}_{-Q}\otimes\MC{A}_n\,.
\end{array}
$$
Explicitly, we have
$$
\begin{array}{l}
\MB{b}^{(+)}_{1^*,\ldots,P^*}(\vec{v})=
{\tsum_{k_1,\ldots,k_P=1}^{n-1}}\MB{f}^{k_1}\otimes\MB{f}^{k_2}\otimes\ldots
\otimes\MB{f}^{k_P}\otimes T^n_{k_1}(v_1)T^n_{k_2}(v_2)\ldots T^n_{k_P}(v_P)\\[6pt]
\MB{b}^{(-)}_{1,\ldots,Q}(\vec{w})=
{\tsum_{r_1,\ldots,r_Q=1}^{n-1}}\MB{e}_{-r_1}\otimes\MB{e}_{-r_2}\otimes\ldots
\otimes\MB{e}_{-r_Q}\otimes
T^{-r_1}_{-n}(w_1)T^{-r_2}_{-n}(w_2)\ldots T^{-r_Q}_{-n}(w_Q)\,.
\end{array}
$$
If we introduce
$\MB{\Phi}\in\MC{V}_1\otimes\cdots\otimes\MC{V}_P\otimes
\MC{V}^*_{-1}\otimes\cdots\otimes\MC{V}^*_{-Q}\otimes\tilde{W}$
$$
\begin{array}{l}
\MB{\Phi}={\tsum_{k_1,\ldots,k_P=1}^{n-1}}\,{\tsum_{r_1,\ldots,r_Q=1}^{n-1}}
\MB{e}_{k_1}\otimes\ldots\otimes\MB{e}_{k_P}\otimes
\MB{f}^{-r_1}\otimes\ldots\otimes\MB{f}^{-r_Q}\otimes
\Phi^{k_1,k_2,\ldots,k_P}_{-r_1,-r_2,\ldots,-r_Q}=\\[6pt]
\hskip20mm=
{\tsum_{\vec{k},\vec{r}}}\MB{e}_{\vec{k}}\otimes\MB{f}^{-\vec{r}}\otimes\Phi^{\vec{k}}_{-\vec{r}}\,,
\end{array}
$$
where
$$
\begin{array}{l}
\Phi^{k_1,k_2,\ldots,k_P}_{-r_1,-r_2,\ldots,-r_Q}=\Phi^{\vec{k}}_{-\vec{r}}\in\tilde{\MC{W}}\,,\\[4pt]
\MB{e}_{\vec{k}}=\MB{e}_{k_1}\otimes\MB{e}_{k_2}\otimes\ldots\otimes\MB{e}_{k_P}\in
\bigl(\MC{V}_+\bigr)^{\otimes P},\\[4pt]
\MB{f}^{-\vec{r}}=\MB{f}^{-r_1}\otimes\MB{f}^{-r_2}\otimes\ldots\otimes\MB{f}^{-r_Q}\in
\bigl(\MC{V}^*_-\bigr)^{\otimes Q},
\end{array}
$$
the assumed shape of the eigenvectors can be written as
$$
\MF{B}(\vec{v},\vec{w})=
\Bigl<\MB{b}^{(+)}_{1^*,\ldots,P^*}(\vec{v})\MB{b}^{(-)}_{1,\ldots,Q}(\vec{w}),
\MB{\Phi}\Bigr>.
$$

\section{Bethe vectors and Bethe condition}
 \label{Bethe}

Our goal is to write the action of the operators $T^n_n(x)$, $T^{-n}_{-n}(x)$,
$\tilde{\MB{T}}^{(+)}$ and $\tilde{\MB{T}}^{(-)}$
on the assumed form of the Bethe vectors using the operators that
act only on $\MB{\Phi}$. These actions are explicitly given in
Lemma 5 of Appendix. Here we list only their consequences.

For $\vec{v}=(v_1,v_2,\ldots,v_P)$ we introduce a set
$\ol{v}=\{v_1,v_2,\ldots,v_P\}$, denote
$$
\begin{array}{ll}
\vec{v}_k=(v_1,\ldots,v_{k-1},v_{k+1},\ldots,v_P)\,,\qquad&
\ol{v}_k=\ol{v}\setminus\{v_k\}\,,\\[4pt]
F(x;\ol{v})={\tprod_{v_k\in\ol{v}}}f(x,v_k)\,,\qquad&
F(\ol{v};x)={\tprod_{v_k\in\ol{v}}}f(v_k,x)\,.
\end{array}
$$
and define
$$
\begin{array}{l}
\wh{\MB{T}}^{(+)}_{0;1,\ldots,P;1^*,\ldots,Q^*}(x;\vec{v};\vec{w})=
\wh{\MB{R}}^{(+,-)}_{0;1^*,\ldots,Q^*}(x;\vec{w})\tilde{\MB{T}}^{(+)}_0(x)
\wh{\MB{R}}^{(+,+)}_{0;1,\ldots,P}(x;\vec{v})=
{\tsum_{i,k=1}^{n-1}}\MB{E}^k_i\otimes\wh{T}^i_k(x;\vec{v};\vec{w})\\[6pt]
\wh{\MB{T}}^{(-)}_{0;1,\ldots,P;1^*,\ldots,Q^*}(x;\vec{v};\vec{w})=
\wh{\MB{R}}^{(-,-)}_{0;1^*,\ldots,Q^*}(x;\vec{w})\tilde{\MB{T}}^{(-)}_0(x)
\wh{\MB{R}}^{(-,+)}_{0;1,\ldots,P}(x;\vec{v})=
{\tsum_{i,k=1}^{n-1}}\MB{E}^{-k}_{-i}\otimes\wh{T}^{-i}_{-k}(x;\vec{v};\vec{w})\,,
\end{array}
$$
where
$$
\begin{array}{l}
\wh{\MB{R}}^{(+,+)}_{0;1,\ldots,P}(x;\vec{v})=
\wh{\MB{R}}^{(+,+)}_{0_+,P_+}(x,v_P)\ldots
\wh{\MB{R}}^{(+,+)}_{0_+,2_+}(x,v_2)
\wh{\MB{R}}^{(+,+)}_{0_+,1_+}(x,v_1)\\[4pt]
\wh{\MB{R}}^{(+,-)}_{0;1^*,\ldots,Q^*}(x;\vec{w})=
\wh{\MB{R}}^{(+,-)}_{0_+,1^*_-}(x,w_1)
\wh{\MB{R}}^{(+,-)}_{0_+,2^*_-}(x,w_2)\ldots
\wh{\MB{R}}^{(+,-)}_{0_+,Q^*_-}(x,w_Q)\\[4pt]
\wh{\MB{R}}^{(-,-)}_{0;1^*,\ldots,Q^*}(x;\vec{w})=
\wh{\MB{R}}^{(-,-)}_{0_-,1^*_-}(x,w_1)
\wh{\MB{R}}^{(-,-)}_{0_-,2^*_-}(x,w_2)\ldots
\wh{\MB{R}}^{(-,-)}_{0_-,Q^*_-}(x,w_Q)\\[4pt]
\wh{\MB{R}}^{(-,+)}_{0;1,\ldots,P}(x;\vec{v})=
\wh{\MB{R}}^{(-,+)}_{0_-,P_+}(x,v_P)\ldots
\wh{\MB{R}}^{(-,+)}_{0_-,2_+}(x,v_2)
\wh{\MB{R}}^{(-,+)}_{0_-,1_+}(x,v_1)\\[4pt]
\wh{\MB{R}}^{(+,+)}_{0_+,1_+}(x,v)=
\dfrac1{f(x,v)}\,\Bigl(\tilde{\MB{I}}_+\otimes\tilde{\MB{I}}_++
g(x,v){\tsum_{i,k=1}^{n-1}}\MB{E}^i_k\otimes\MB{E}^k_i\Bigr)\\[4pt]
\wh{\MB{R}}^{(+,-)}_{0_+,1^*_-}(x,w)=
\tilde{\MB{I}}_+\otimes\tilde{\MB{I}}^*_--
\tilde{h}(w,x){\tsum_{r,s=1}^{n-1}}\MB{E}^r_s\otimes\MB{F}^{-r}_{-s}\\[4pt]
\tilde{\MB{R}}^{(-,+)}_{0_-,1_+}(x,v)=
\tilde{\MB{I}}_-\otimes\tilde{\MB{I}}_+-
\tilde{h}(x,v){\tsum_{i,k=1}^{n-1}}\MB{E}^{-i}_{-k}\otimes\MB{E}^i_k\\[4pt]
\wh{\MB{R}}^{(-,-)}_{0_-,1^*_-}(x,w)=
\dfrac1{f(w,x)}\Bigl(\tilde{\MB{I}}_{0_-}\otimes\tilde{\MB{I}}^*_{1_-}+
g(w,x){\tsum_{r,s=1}^{n-1}}\MB{E}^{-r}_{-s}\otimes\MB{F}^{-s}_{-r}\Bigr).
\end{array}
$$

The main results of this paper are the following three Theorems.

\medskip

\underbar{\textbf{Theorem 2.}} Let
$\MB{\Phi}={\tsum_{\vec{k},\vec{r}}}
\MB{e}_{\vec{k}}\otimes\MB{f}^{-\vec{r}}\otimes\Phi^{\vec{k}}_{-\vec{r}}$,
where $\Phi^{\vec{k}}_{-\vec{r}}\in\tilde{\MC{W}}$ is an common eigenvector
of the operators
$$
\begin{array}{l}
\wh{H}^{(+)}(x;\vec{v};\vec{w})=
\Tr_0\bigl(\wh{\MB{T}}^{(+)}_{0;1,\ldots,P;1^*,\ldots,Q^*}(x;\vec{v};\vec{w})\bigr)=
{\tsum_{i=1}^{n-1}}\wh{T}^i_i(x;\vec{v};\vec{w})\\[6pt]
\wh{H}^{(-)}(x;\vec{v};\vec{w})=
\Tr_0\bigl(\wh{\MB{T}}^{(-)}_{0;1,\ldots,P;1^*,\ldots,Q^*}(x;\vec{v};\vec{w})\bigr)=
{\tsum_{i=1}^{n-1}}\wh{T}^{-i}_{-i}(x;\vec{v};\vec{w})
\end{array}
$$
with eigenvalues $\mu^{(+)}(x;\vec{v};\vec{w})$ and
$\mu^{(-)}(x;\vec{v};\vec{w})$. If the Bethe conditions
\begin{equation}
 \label{BP-n}
\begin{array}{l}
\lambda_n(v_\ell)F(\ol{v}_\ell;v_\ell)F(\ol{w};v_\ell-n+1+\eta)=
\mu^{(+)}(v_\ell;\vec{v};\vec{w})F(v_\ell;\ol{v}_\ell)\\[6pt]
\lambda_{-n}(w_s)F(w_s;\ol{w}_s)F(w_s+n-1-\eta;\ol{v})=
\mu^{(-)}(w_s;\vec{v};\vec{w})F(\ol{w}_s;w_s)
\end{array}
\end{equation}
are fulfilled for any $v_\ell\in\ol{v}$ and $w_s\in\ol{w}$, the vector
$$
\MF{B}(\vec{v};\vec{w})=
\Bigl<\MB{b}^{(+)}_{1^*,\ldots,P^*}(\vec{v})\MB{b}^{(-)}_{1,\ldots,Q}(\vec{w}),\MB{\Phi}\Bigr>
$$
is a common eigenvector of the operators $H^{(+)}(x)$ and $H^{(-)}(x)$
with eigenvalues
$$
\begin{array}{l}
E^{(+)}(x;\vec{v};\vec{w})=
\lambda_n(x)F(\ol{v};x)F(\ol{w};x-n+1+\eta)+
\mu^{(+)}(x;\vec{v},\vec{w})F(x;\ol{v})\\[6pt]
E^{(-)}(x;\vec{v};\vec{w})=
\lambda_{-n}(x)F(x;\ol{w})F(x+n-1-\eta;\ol{v})+
\mu^{(-)}(x;\vec{v};\vec{w})F(\ol{w};x)\,.
\end{array}
$$

\textsc{Proof:} According to Proposition 1, we have
$T^n_n(x)\MB{\Phi}=\lambda_n(x)\MB{\Phi}$ and
$T^{-n}_{-n}(x)\MB{\Phi}=\lambda_{-n}(x)\MB{\Phi}$.

Using Lemma 5 and the relations $\Tr_0\wh{\R}^{(+,-)}_{0_+,s^*_-}=\wh{\MB{I}}_{s^*_-}$
and $\Tr_0\wh{\R}^{(-,+)}_{0_-,\ell_+}=\wh{\MB{I}}_{\ell_-}$ we
obtain
$$
\begin{array}{l}
H^{(+)}(x)\Bigl<\MB{b}^{(+)}_{1^*,\ldots,P^*}(\vec{v})
\MB{b}^{(-)}_{1,\ldots,Q}(\vec{w}),\MB{\Phi}\Bigr>=\\[6pt]
\hskip5mm=
\Bigl(\lambda_n(x)F(\ol{v};x)F(\ol{w};x-n+1+\eta)+
\mu^{(+)}(x;\vec{v},\vec{w})F(x;\ol{v})\Bigr)\\[4pt]
\hskip60mm
\Bigl<\MB{b}^{(+)}_{1^*,\ldots,P^*}(\vec{v})\MB{b}^{(-)}_{1,\ldots,Q}(\vec{w}),
\MB{\Phi}\Bigr>-\\[4pt]
\hskip10mm-
{\tsum_{v_\ell\in\ol{v}}}g(v_\ell,x)
\Bigl(\lambda_n(v_\ell)F(\ol{v}_\ell;v_\ell)F(\ol{w};v_\ell-n+1+\eta)-
\mu^{(+)}(v_\ell;\vec{v};\vec{w})F(v_\ell;\ol{v}_\ell)\Bigr)\\[4pt]
\hskip60mm
\Bigl<\MB{b}^{(+)}_{\ell^*;1^*,\ldots,P^*}(x;\vec{v}_\ell)
\MB{b}^{(-)}_{1,\ldots,Q}(\vec{w}),
\wh{\MB{R}}^{(+,+)}_{1,\ldots,\ell}(\vec{v})\MB{\Phi}\Bigr>-\\[4pt]
\hskip10mm-
{\tsum_{w_s\in\ol{w}}}\tilde{h}(w_s,x)
\Bigl(\lambda_{-n}(w_s)F(w_s;\ol{w}_s)F(w_s+n-1-\eta;\ol{v})-
\mu^{(-)}(w_s;\vec{v};\vec{w})F(\ol{w}_s;w_s)\Bigr)\\[4pt]
\hskip60mm
\Bigl<\MB{b}^{(+)}_{s^*}(x)\MB{b}^{(+)}_{1^*,\ldots,P^*}(\vec{v})
\MB{b}^{(-)}_{1,\ldots\wh{s},\ldots,Q}(\vec{w}_s),
\wh{\BB{P}}^{(+,-)}_{s_+,s^*_-}\wh{\MB{R}}^{(-,-)}_{1^*,\ldots,s^*}(\vec{w})
\MB{\Phi}\Bigr>\\[9pt]
H^{(-)}(x)\Bigl<\MB{b}^{(+)}_{1^*,\ldots,P^*}(\vec{v})
\MB{b}^{(-)}_{1,\ldots,Q}(\vec{w}),\MB{\Phi}\Bigr>=\\
\end{array}
$$
This immediately proves the statement of Theorem 2. \qed

\bigskip

\underbar{\textbf{Theorem 3.}} The operators
$\wh{T}^i_k(x;\vec{v};\vec{w})$ and
$\wh{T}^{-i}_{-k}(x;\vec{v};\vec{w})$ are for any $\vec{v}$
and $\vec{w}$ generators of the RTT--algebra of $\MC{A}_{n-1}$ type.

\smallskip

\textsc{Proof:}
We have to show that for any $\vec{v}$, $\vec{w}$
and $\epsilon,\,\epsilon'=\pm$ the relation
$$
\begin{array}{l}
\tilde{\MB{R}}^{(\epsilon,\epsilon')}_{0,0'}(x,y)
\wh{\MB{T}}^{(\epsilon)}_{0;1,\ldots,P;1^*,\ldots,Q^*}(x;\vec{v};\vec{w})
\wh{\MB{T}}^{(\epsilon')}_{0';1,\ldots,P;1^*,\ldots,Q^*}(y;\vec{v};\vec{w})=\\[4pt]
\hskip40mm=
\wh{\MB{T}}^{(\epsilon')}_{0';1,\ldots,P;1^*,\ldots,Q^*}(y;\vec{v};\vec{w})
\wh{\MB{T}}^{(\epsilon)}_{0;1,\ldots,P;1^*,\ldots,Q^*}(x;\vec{v};\vec{w})
\tilde{\MB{R}}^{(\epsilon,\epsilon')}_{0,0'}(x,y)\,.
\end{array}
$$
is valid. Since
$$
\wh{\MB{T}}^{(\epsilon)}_{0;1,\ldots,P;1^*,\ldots,Q^*}(x;\vec{v};\vec{w})=
\wh{\MB{R}}^{(\epsilon,-)}_{0;1^*,\ldots,Q^*}(x;\vec{w})
\tilde{\MB{T}}^{(\epsilon)}_0(x)
\wh{\MB{R}}^{(\epsilon,+)}_{0;1,\ldots,P}(x;\vec{v})
$$
and $\tilde{\MB{T}}^{(\epsilon)}_0(x)$ satisfies the RTT--equation, it
is enough to show that
$$
\begin{array}{l}
\tilde{\MB{R}}^{(\epsilon,\epsilon')}_{0,0'}(x,y)
\wh{\MB{R}}^{(\epsilon,-)}_{0;1^*,\ldots,Q^*}(x;\vec{w})
\wh{\MB{R}}^{(\epsilon',-)}_{0';1^*,\ldots,Q^*}(y;\vec{w})=
\wh{\MB{R}}^{(\epsilon',-)}_{0';1^*,\ldots,Q^*}(y;\vec{w})
\wh{\MB{R}}^{(\epsilon,-)}_{0;1^*,\ldots,Q^*}(x;\vec{w})
\tilde{\MB{R}}^{(\epsilon,\epsilon;)}_{0,0'}(x,y)\\[4pt]
\tilde{\MB{R}}^{(\epsilon,\epsilon')}_{0,0'}(x,y)
\wh{\MB{R}}^{(\epsilon,+)}_{0;1,\ldots,P}(x;\vec{v})
\wh{\MB{R}}^{(\epsilon',+)}_{0';1,\ldots,P}(y;\vec{v})=
\wh{\MB{R}}^{(\epsilon',+)}_{0';1,\ldots,P}(y;\vec{v})
\wh{\MB{R}}^{(\epsilon,+)}_{0;1,\ldots,P}(x;\vec{v})
\tilde{\MB{R}}^{(\epsilon,\epsilon')}_{0,0'}(x,y)\,.
\end{array}
$$
hold. According to the definitions, we have
$$
\begin{array}{l}
\wh{\MB{R}}^{(\epsilon,-)}_{0;1^*,\ldots,Q^*}(x;\vec{w})=
\wh{\MB{R}}^{(\epsilon,-)}_{0_\epsilon,1^*_-}(x,w_1)
\wh{\MB{R}}^{(\epsilon,-)}_{0_\epsilon,2^*_-}(x,w_2)\ldots
\wh{\MB{R}}^{(\epsilon,-)}_{0_\epsilon,Q^*_-}(x,w_Q)\\[4pt]
\wh{\MB{R}}^{(\epsilon,+)}_{0;1,\ldots,P}(x;\vec{v})=
\wh{\MB{R}}^{(\epsilon,+)}_{0_\epsilon,P_+}(x,v_P)\ldots
\wh{\MB{R}}^{(\epsilon,+)}_{0_\epsilon,2_+}(x,v_2)
\wh{\MB{R}}^{(\epsilon,+)}_{0_\epsilon,1_+}(x,v_1)\,.
\end{array}
$$
and a theorem then follows from the Yang--Baxter equations
$$
\begin{array}{l}
\tilde{\MB{R}}^{(\epsilon,\epsilon')}_{0,0'}(x,y)
\wh{\MB{R}}^{(\epsilon,-)}_{0_\epsilon,1^*_-}(x,w)
\wh{\MB{R}}^{(\epsilon',-)}_{0'_{\epsilon'},1^*_-}(y,w)=
\wh{\MB{R}}^{(\epsilon',-)}_{0'_{\epsilon'},1^*_-}(y,w)
\wh{\MB{R}}^{(\epsilon,-)}_{0_\epsilon,1^*_-}(x,w)
\tilde{\MB{R}}^{(\epsilon,\epsilon')}_{0,0'}(x,y)\,,\\[4pt]
\tilde{\MB{R}}^{(\epsilon,\epsilon')}_{0,0'}(x,y)
\wh{\MB{R}}^{(\epsilon,+)}_{0_\epsilon,1_+}(x,v)
\wh{\MB{R}}^{(\epsilon',+)}_{0'_{\epsilon'},1_+}(y,v)=
\wh{\MB{R}}^{(\epsilon',+)}_{0'_{\epsilon'},1_+}(y,v)
\wh{\MB{R}}^{(\epsilon,+)}_{0_\epsilon,1_+}(x,v)
\tilde{\MB{R}}^{(\epsilon,\epsilon')}_{0,0'}(x,y)\,.
\end{array}
$$
\qed

\bigskip

The following Theorem shows that
$$
\wh{\MB{\Omega}}=
\underbrace{\MB{e}_{n-1}\otimes\ldots\otimes\MB{e}_{n-1}}_{P\times}\otimes
\underbrace{\MB{f}^{-n+1}\otimes\ldots\otimes\MB{f}^{-n+1}}_{Q\times}\otimes
\omega
$$
is a vacuum vector for the representation of the RTT--algebra
$\MC{A}_{n-1}$, which is generated by $\wh{T}^i_k(x;\vec{v};\vec{w})$
and $\wh{T}^{-i}_{-k}(x;\vec{v};\vec{w})$.

\medskip

\underbar{\textbf{Theorem 4.}} For the vector $\wh{\MB{\Omega}}$ and
$i,\,k=1,\,\ldots,\,n-1$
$$
\begin{array}{ll}
\wh{T}^i_k(x;\vec{v},\vec{w})\wh{\MB{\Omega}}=0\quad \MR{for}\quad
i<k\,,\qquad&
\wh{T}^{-i}_{-k}(x;\vec{v};\vec{w})\wh{\MB{\Omega}}=0\quad
\MR{for}\quad k<i\\[4pt]
\wh{T}^i_i(x;\vec{v};\vec{w})\wh{\MB{\Omega}}=
\nu_i(x;\vec{v};\vec{w})\wh{\MB{\Omega}}\qquad&
\wh{T}^{-i}_{-i}(x;\vec{v};\vec{w})\wh{\MB{\Omega}}=
\nu_{-i}(x;\vec{v};\vec{w})\wh{\MB{\Omega}}
\end{array}
$$
where
$$
\begin{array}{lcl}
\nu_i(x;\vec{v};\vec{w})=
\lambda_i(x)F(\ol{v};x+1)&\quad\MR{for}\quad&
1\leq i<n-1\\[4pt]
\nu_{n-1}(x;\vec{v};\vec{w})=
\lambda_{n-1}(x)F(x-n+1+\eta;\ol{w})\quad&\\[4pt]
\nu_{-i}(x;\vec{v};\vec{w})=
\lambda_{-i}(x)F(x-1;\ol{w})&\quad\MR{for}\quad&
1\leq i<n-1\\[4pt]
\nu_{-n+1}(x;\vec{v};\vec{w})=
\lambda_{-n+1}(x)F(\ol{v};x+n-1-\eta)\quad&
\end{array}
$$
are valid.

\textsc{Proof:} If we write
$$
\begin{array}{ll}
\wh{\MB{R}}^{(+,+)}_{0_+,1_+}(x,v)=
{\tsum_{a,b,p,q=1}^{n-1}}R^{a,p}_{b,q}(x,v)\MB{E}^b_a\otimes\MB{E}^q_p&
\qquad
R^{a,p}_{b,q}(x,v)=\dfrac{\delta^a_b\delta^p_q+g(x,v)\delta^a_q\delta^p_b}{f(x,v)}\\[4pt]
\wh{\MB{R}}^{(+,-)}_{0_+,1^*_-}(x,w)=
{\tsum_{c,d,r,s=1}^{n-1}}R^{c,-r}_{d,-s}(x,w)\MB{E}^d_c\otimes\MB{F}^{-s}_{-r}&
\qquad
R^{c,-r}_{d,-s}(x,w)=\delta^c_d\delta^r_s-\tilde{h}(w,x)\delta^{c,r}\delta_{d,s}\,,
\end{array}
$$
$$
\begin{array}{ll}
\wh{\MB{R}}^{(-,+)}_{0_-,1_+}(x,v)=
{\tsum_{a,b,p,q=1}^{n-1}}R^{-a,p}_{-b,q}(x,v)\MB{E}^{-b}_{-a}\otimes\MB{E}^q_p&
\qquad
R^{-a,p}_{-b,q}(x,v)=\delta^a_b\delta^p_q-\tilde{h}(x,v)\delta^{a,p}\delta_{b,q}\\[4pt]
\wh{\MB{R}}^{(-,-)}_{0_-,1^*_-}(x,w)=
{\tsum_{c,d,r,s=1}^{n-1}}R^{-c,-r}_{-d,-s}(x,w)\MB{E}^{-d}_{-c}\otimes\MB{F}^{-s}_{-r}&
\qquad
R^{-c,-r}_{-d,-s}(x,w)=\dfrac{\delta^c_d\delta^r_s+g(w,x)\delta^c_s\delta^r_d}{f(w,x)}\,,
\end{array}
$$
we obtain
$$
\begin{array}{l}
\wh{T}^i_k(x;\vec{v};\vec{w})\wh{\MB{\Omega}}=\\[6pt]
\hskip10mm= R^{i,-n+1}_{d_1,-s_1}(x,w_1)
R^{d_1,-n+1}_{d_2,-s_2}(x,w_2) \ldots
R^{d_{Q-2},-n+1}_{d_{Q-1},-s_{Q-1}}(x,w_{Q-1})
R^{d_{Q-1},-n+1}_{d_Q,-s_Q}(x,w_Q)\\[4pt]
\hskip20mm R^{a_P,p_P}_{a_{P-1},n-1}(x,v_P)
R^{a_{P-1},p_{P-1}}_{a_{P-2},n-1}(x,v_{P-1})\ldots
R^{a_3,p_2}_{a_1,n-1}(x,v_2)
R^{a_1,p_1}_{k,n-1}(x,v_1)\\[6pt]
\hskip30mm \MB{e}_{p_1}\otimes \MB{e}_{p_2}\otimes\ldots\otimes
\MB{e}_{p_{P-1}}\otimes
\MB{e}_{p_P}\otimes\\[4pt]
\hskip40mm\otimes \MB{f}^{-s_1}\otimes
\MB{f}^{-s_2}\otimes\ldots\otimes \MB{f}^{-s_{Q-1}}\otimes
\MB{f}^{-s_Q}\otimes
T^{d_Q}_{a_P}(x)\omega\\[9pt]
\wh{T}^{-i}_{-k}(x;\vec{v};\vec{w})
\wh{\MB{\Omega}}=\\[6pt]
\hskip10mm= R^{-i,-n+1}_{-d_1,-s_1}(x,w_1)
R^{-d_1,-n+1}_{-d_2,-s_2}(x,w_2) \ldots
R^{-d_{Q-2},-n+1}_{-d_{Q-1},-s_{Q-1}}(x,w_{Q-1})
R^{-d_{Q-1},-n+1}_{-d_Q,-s_Q}(x,w_Q)\\[4pt]
\hskip20mm R^{-a_P,p_P}_{-a_{P-1},n-1}(x,v_P)
R^{-a_{P-1},p_{P-1}}_{-a_{P-2},n-1}(x,v_{P-1})\ldots
R^{-a_2,p_2}_{-a_1,n-1}(x,v_2)
R^{-a_1,p_1}_{-k,n-1}(x,v_1)\\[4pt]
\hskip30mm \MB{e}_{p_1}\otimes \MB{e}_{p_2}\otimes\ldots\otimes
\MB{e}_{p_{P-1}}\otimes
\MB{e}_{p_P}\otimes\\[4pt]
\hskip40mm\otimes \MB{f}^{-s_1}\otimes
\MB{f}^{-s_2}\otimes\ldots\otimes \MB{f}^{-s_{Q-1}}\otimes
\MB{f}^{-s_Q}\otimes T^{-d_Q}_{-a_P}(x)\omega
\end{array}
$$

Since $R^{i,-n+1}_{d,-s}(x,w)=\delta^i_d\delta^{n-1}_s$  for $1\leq
i<n-1$, we have
$$
\begin{array}{l}
\wh{T}^i_k(x;\vec{v};\vec{w}) \wh{\MB{\Omega}}=
R^{a_P,p_P}_{a_{P-1},n-1}(x,v_P)
R^{a_{P-1},p_{P-1}}_{a_{P-2},n-1}(x,v_{P-1})\ldots
R^{a_3,p_2}_{a_1,n-1}(x,v_2)
R^{a_1,p_1}_{k,n-1}(x,v_1)\\[6pt]
\hskip20mm \MB{e}_{p_1}\otimes \MB{e}_{p_2}\otimes\ldots\otimes
\MB{e}_{p_{P-1}}\otimes
\MB{e}_{p_P}\otimes\\[4pt]
\hskip30mm\otimes \MB{f}^{-n+1}\otimes
\MB{f}^{-n+1}\otimes\ldots\otimes \MB{f}^{-n+1}\otimes
\MB{f}^{-n+1}\otimes T^i_{a_P}(x)\omega
\end{array}
$$
As $T^i_{a_P}(x)\omega=0$ for $a_P>i$, this expression is
nonzero only for $a_P\leq i<n-1$. In this case
$R^{a_P,p_P}_{a_{P-1},n-1}(x,v_P)=
\dfrac1{f(x,v_P)}\,\delta^{a_P}_{a_{P-1}}\delta^{p_P}_{n-1}$.
Therefore, we have
$$
\begin{array}{l}
\wh{T}^i_k(x;\vec{v};\vec{w}) \wh{\MB{\Omega}}= \dfrac1{F(x;\ol{v})}
\MB{e}_{n-1}\otimes \MB{e}_{n-1}\otimes\ldots\otimes
\MB{e}_{n-1}\otimes
\MB{e}_{n-1}\otimes\\[4pt]
\hskip30mm\otimes \MB{f}^{-n+1}\otimes
\MB{f}^{-n+1}\otimes\ldots\otimes \MB{f}^{-n+1}\otimes
\MB{f}^{-n+1}\otimes T^i_k(x)\omega
\end{array}
$$
and so
$$
\wh{T}^i_k(x;\vec{v};\vec{w})\wh{\MB{\Omega}}=0 \quad\MR{for}\quad
k>i\,,\qquad \wh{T}^i_i(x;\vec{v};\vec{w})\wh{\MB{\Omega}}=
\dfrac{\lambda_i(x)}{F(x;\ol{v})}\,\wh{\MB{\Omega}}=
\lambda_i(x)F(\ol{v};x+1)\wh{\MB{\Omega}}\,.
$$

If $i=k=n-1$ we have
$$
\begin{array}{l}
\wh{T}^{n-1}_{n-1}(x;\vec{v};\vec{w})
\wh{\MB{\Omega}}=\\[6pt]
\hskip10mm= R^{n-1,-n+1}_{d_1,-s_1}(x,w_1)
R^{d_1,-n+1}_{d_2,-s_2}(x,w_2) \ldots
R^{d_{Q-2},-n+1}_{d_{Q-1},-s_{Q-1}}(x,w_{Q-1})
R^{d_{Q-1},-n+1}_{d_Q,-s_Q}(x,w_Q)\\[4pt]
\hskip20mm R^{a_P,p_P}_{a_{P-1},n-1}(x,v_P)
R^{a_{P-1},p_{P-1}}_{a_{P-2},n-1}(x,v_{P-1})\ldots
R^{a_3,p_2}_{a_1,n-1}(x,v_2)
R^{a_1,p_1}_{n-1,n-1}(x,v_1)\\[6pt]
\hskip30mm \MB{e}_{p_1}\otimes \MB{e}_{p_2}\otimes\ldots\otimes
\MB{e}_{p_{P-1}}\otimes
\MB{e}_{p_P}\otimes\\[4pt]
\hskip40mm\otimes \MB{f}^{-s_1}\otimes
\MB{f}^{-s_2}\otimes\ldots\otimes \MB{f}^{-s_{Q-1}}\otimes
\MB{f}^{-s_Q}\otimes T^{d_Q}_{a_P}(x)\omega
\end{array}
$$
Since $R^{a,p}_{n-1,n-1}(x,v)=\delta^a_{n-1}\delta^p_{n-1}$
we obtain
$$
\begin{array}{l}
\wh{T}^{n-1}_{n-1}(x;\vec{v};\vec{w})
\wh{\MB{\Omega}}=\\[6pt]
\hskip10mm= R^{n-1,-n+1}_{d_1,-s_1}(x,w_1)
R^{d_1,-n+1}_{d_2,-s_2}(x,w_2) \ldots
R^{d_{Q-2},-n+1}_{d_{Q-1},-s_{Q-1}}(x,w_{Q-1})
R^{d_{Q-1},-n+1}_{d_Q,-s_Q}(x,w_Q)\\[4pt]
\hskip20mm \MB{e}_{n-1}\otimes \MB{e}_{n-1}\otimes\ldots\otimes
\MB{e}_{n-1}\otimes
\MB{e}_{n-1}\otimes\\[4pt]
\hskip30mm\otimes \MB{f}^{-s_1}\otimes
\MB{f}^{-s_2}\otimes\ldots\otimes \MB{f}^{-s_{Q-1}}\otimes
\MB{f}^{-s_Q}\otimes T^{d_Q}_{n-1}(x)\omega\,.
\end{array}
$$
The conditions $T^{d_Q}_{n-1}(x)\omega=0$ for $d_Q<n-1$ and
$T^{n-1}_{n-1}(x)\omega=\lambda_{n-1}(x)\omega$
lead to the equations
$$
\begin{array}{l}
\wh{T}^{n-1}_{n-1}(x;\vec{v};\vec{w})
\wh{\MB{\Omega}}=\\[6pt]
\hskip7mm= \lambda_{n-1}(x) R^{n-1,-n+1}_{d_1,-s_1}(x,w_1)
R^{d_1,-n+1}_{d_2,-s_2}(x,w_2) \ldots
R^{d_{Q-2},-n+1}_{d_{Q-1},-s_{Q-1}}(x,w_{Q-1})
R^{d_{Q-1},-n+1}_{n-1,-s_Q}(x,w_Q)\\[4pt]
\hskip20mm \MB{e}_{n-1}\otimes \MB{e}_{n-1}\otimes\ldots\otimes
\MB{e}_{n-1}\otimes
\MB{e}_{n-1}\otimes\\[4pt]
\hskip30mm\otimes \MB{f}^{-s_1}\otimes
\MB{f}^{-s_2}\otimes\ldots\otimes \MB{f}^{-s_{Q-1}}\otimes
\MB{f}^{-s_Q}\otimes\omega
\end{array}
$$
However,
$R^{d,-n+1}_{n-1,-s}(x,w)=\bigl(1-\tilde{h}(w,x)\bigr)\delta^d_{n-1}\delta^{n-1}_s=
f(x-n+1+\eta,w)\delta^d_{n-1}\delta^{n-1}_s$, and so
$$
\wh{T}^{n-1}_{n-1}(x;\vec{v};\vec{w}) \wh{\MB{\Omega}}=
\lambda_{n-1}(x)F(x-n+1+\eta;\ol{w})\wh{\MB{\Omega}}.
$$

\bigskip

Since $R^{-a,p}_{-k,n-1}(x,v)=\delta^a_k\delta^p_{n-1}$ for $1\leq k<n-1$
we have
$$
\begin{array}{l}
\wh{T}^{-i}_{-k}(x;\vec{v};\vec{w})
\wh{\MB{\Omega}}=\\[6pt]
\hskip10mm= R^{-i,-n+1}_{-d_1,-s_1}(x,w_1)
R^{-d_1,-n+1}_{-d_2,-s_2}(x,w_2) \ldots
R^{-d_{Q-2},-n+1}_{-d_{Q-1},-s_{Q-1}}(x,w_{Q-1})
R^{-d_{Q-1},-n+1}_{-d_Q,-s_Q}(x,w_Q)\\[4pt]
\hskip20mm \MB{e}_{n-1}\otimes \MB{e}_{n-1}\otimes\ldots\otimes
\MB{e}_{n-1}\otimes
\MB{e}_{n-1}\otimes\\[4pt]
\hskip30mm\otimes \MB{f}^{-s_1}\otimes
\MB{f}^{-s_2}\otimes\ldots\otimes \MB{f}^{-s_{Q-1}}\otimes
\MB{f}^{-s_Q}\otimes T^{-d_Q}_{-k}(x)\omega
\end{array}.
$$
For $k<d_Q$ the relation $T^{-d_Q}_{-k}(x)\omega=0$
holds. Therefore, this expression is nonzero for
$1\leq d_Q\leq k<n-1$ only. But in this case
$R^{-d_{Q-1},-n+1}_{-d_Q,-s_Q}(x,w_Q)=
\dfrac1{f(w_Q,x)}\,\delta^{d_{Q-1}}_{\delta_Q}\delta^{n-1}_{s_Q}$.
By repeatedly using this relationship, we get
$$
\begin{array}{l}
\wh{T}^{-i}_{-k}(x;\vec{v};\vec{w}) \wh{\MB{\Omega}}=
\dfrac1{F(\ol{w};x)}\,
\MB{e}_{n-1}\otimes\MB{e}_{n-1}\otimes\ldots\otimes\MB{e}_{n-1}\otimes\MB{e}_{n-1}\otimes\\[6pt]
\hskip30mm\otimes
\MB{f}^{-n+1}\otimes\MB{f}^{-n+1}\otimes\ldots\otimes\MB{f}^{-n+1}\otimes\MB{f}^{-n+1}\otimes
T^{-i}_{-k}(x)\omega
\end{array}
$$
The relations $T^{-i}_{-k}(x)\omega=0$ for $k<i$ and
$T^{-i}_{-i}(x)\omega=\lambda_{-i}(x)\omega$ lead to the equations
$\wh{T}^{-i}_{-k}(x;\vec{v};\vec{w})\wh{\MB{\Omega}}=0$ for $k<i$ and
$$
\wh{T}^{-i}_{-i}(x;\vec{v};\vec{w}) \wh{\MB{\Omega}}=
\dfrac{\lambda_{-i}(x)}{F(\ol{w};x)}\,\wh{\MB{\Omega}}=
\lambda_{-i}(x)F(x-1;\ol{w})\wh{\MB{\Omega}}\,.
$$

\bigskip

For $i=k=n-1$ we have
$$
\begin{array}{l}
\wh{T}^{-n+1}_{-n+1}(x;\vec{v};\vec{w})
\wh{\MB{\Omega}}=\\[6pt]
\hskip10mm= R^{-n+1,-n+1}_{-d_1,-s_1}(x,w_1)
R^{-d_1,-n+1}_{-d_2,-s_2}(x,w_2) \ldots
R^{-d_{Q-2},-n+1}_{-d_{Q-1},-s_{Q-1}}(x,w_{Q-1})
R^{-d_{Q-1},-n+1}_{-d_Q,-s_Q}(x,w_Q)\\[4pt]
\hskip20mm R^{-a_P,p_P}_{-a_{P-1},n-1}(x,v_P)
R^{-a_{P-1},p_{P-1}}_{-a_{P-2},n-1}(x,v_{P-1})\ldots
R^{-a_2,p_2}_{-a_1,n-1}(x,v_2)
R^{-a_1,p_1}_{-n+1,n-1}(x,v_1)\\[4pt]
\hskip30mm \MB{e}_{p_1}\otimes \MB{e}_{p_2}\otimes\ldots\otimes
\MB{e}_{p_{P-1}}\otimes
\MB{e}_{p_P}\otimes\\[4pt]
\hskip40mm\otimes \MB{f}^{-s_1}\otimes
\MB{f}^{-s_2}\otimes\ldots\otimes \MB{f}^{-s_{Q-1}}\otimes
\MB{f}^{-s_Q}\otimes T^{-d_Q}_{-a_P}(x)\omega
\end{array}
$$
Since $R^{-n+1,-n+1}_{-d,-s}(x,w)=\delta^{n-1}_d\delta^{n-1}_s$,
we obtain
$$
\begin{array}{l}
\wh{T}^{-n+1}_{-n+1}(x;\vec{v};\vec{w})
\wh{\MB{\Omega}}=\\[6pt]
\hskip10mm= R^{-a_P,p_P}_{-a_{P-1},n-1}(x,v_P)
R^{-a_{P-1},p_{P-1}}_{-a_{P-2},n-1}(x,v_{P-1})\ldots
R^{-a_2,p_2}_{-a_1,n-1}(x,v_2)
R^{-a_1,p_1}_{-n+1,n-1}(x,v_1)\\[4pt]
\hskip20mm \MB{e}_{p_1}\otimes \MB{e}_{p_2}\otimes\ldots\otimes
\MB{e}_{p_{P-1}}\otimes
\MB{e}_{p_P}\otimes\\[4pt]
\hskip30mm\otimes \MB{f}^{-n+1}\otimes
\MB{f}^{-n+1}\otimes\ldots\otimes \MB{f}^{-n+1}\otimes
\MB{f}^{-n+1}\otimes T^{-n+1}_{-a_P}(x)\omega
\end{array}
$$
The conditions $T^{-n+1}_{-a_P}(x)\omega=0$ for $a_P<n-1$ and
$T^{-n+1}_{-n+1}(x)\omega=\lambda_{-n+1}(x)\omega$ lead to
$$
\begin{array}{l}
\wh{T}^{-n+1}_{-n+1}(x;\vec{v};\vec{w})
\wh{\MB{\Omega}}=\\[6pt]
\hskip9mm= \lambda_{-n+1}(x) R^{-n+1,p_P}_{-a_{P-1},n-1}(x,v_P)
R^{-a_{P-1},p_{P-1}}_{-a_{P-2},n-1}(x,v_{P-1})\ldots
R^{-a_2,p_2}_{-a_1,n-1}(x,v_2)
R^{-a_1,p_1}_{-n+1,n-1}(x,v_1)\\[4pt]
\hskip20mm \MB{e}_{p_1}\otimes \MB{e}_{p_2}\otimes\ldots\otimes
\MB{e}_{p_{P-1}}\otimes
\MB{e}_{p_P}\otimes\\[4pt]
\hskip30mm\otimes \MB{f}^{-n+1}\otimes
\MB{f}^{-n+1}\otimes\ldots\otimes \MB{f}^{-n+1}\otimes
\MB{f}^{-n+1}\otimes\omega
\end{array}
$$
However,
$R^{-n+1,p}_{-a,n-1}(x,v)=\bigl(1-\tilde{h}(x,v)\bigr)\delta^{n-1}_a\delta^p_{n-1}=
f(v,x+n-1-\eta)\delta^{n-1}_a\delta^p_{n-1}$, and so
$$
\wh{T}^{-n+1}_{-n+1}(x;\vec{v};\vec{w}) \wh{\MB{\Omega}}=
\lambda_{-n+1}(x)F(\ol{v};x+n-1-\eta)\wh{\MB{\Omega}}
$$
\qed

\bigskip

These three theorems show that to find the Bethe vectors
$\MF{B}(\vec{v};\vec{w})$ for the RTT--algebra $\MC{A}_n$, it is
sufficient to find the Bethe vectors for the RTT--algebra $\MC{A}_{n-1}$
that is generated by the operators $\wh{T}^i_k(x;\vec{v};\vec{w})$,
$\wh{T}^{-i}_{-k}(x;\vec{v};\vec{w})$, where $i,k=1,\ldots,n-1$,
and that has a vacuum vector $\wh{\MB{\Omega}}$.

\section{Conclusion}
 \label{zaver}

The paper describes the construction of eigenvectors for the representations
of the RTT--algebra $\MC{A}_n$ by using the highest weight vectors
for the representation of the RTT--algebra $\MC{A}_{n-1}$.
We meet these RTT--algebras \cite{BN-SP, BN-SO} while studying the
algebraic Bethe ansatz for the RTT--algebras of $\MR{sp}(2n)$ and
$\MR{o}(2n)$ types.

In the special cases, when $\ol{v}$ or $\ol{w}$ is an empty set, our
construction is known as the algebraic nested Bethe ansatz, which
was formulated in \cite{KR83}. So our construction of the Bethe
vectors is a generalization of the algebraic nested Bethe ansatz to
the RTT--algebra of $\MC{A}_n$ type.

For the RTT--algebra of $\MC{A}_2$ type we get from theorems 2, 3 and 4 the Bethe
vectors
$$
\MF{B}_2(\vec{v};\vec{w})= T^2_1(\vec{v})T^{-1}_{-2}(\vec{w})\omega
$$
and the Bethe conditions
$$
\begin{array}{l}
\lambda_2(v_\ell)F(\ol{v}_\ell;v_\ell)F(\ol{w};v_\ell-1+\eta)=
\lambda_1(v_\ell)F(v_\ell-1+\eta;\ol{w})F(v_\ell;\ol{v}_\ell)\\[6pt]
\lambda_{-2}(w_s)F(w_s;\ol{w}_s)F(w_s+1-\eta;\ol{v})=
\lambda_{-1}(w_s)F(\ol{v};w_s+1-\eta)\,F(\ol{w}_s;w_s),
\end{array}
$$
which we found for this algebra and $\nu=-1$ in \cite{BN-SP4}.

For higher $n$ it is possible by means of Theorems 2, 3 and 4
step-by-step to decrease value $n$ and thus obtain an explicit form of the
Bethe vectors. For the RTT--algebra of $\MR{gl}(n)$ type
this procedure leads to trace-formula \cite{MTV-2006}.
We intend to publish a similar explicit form of the Bethe vectors
for the RTT--algebras $\MC{A}_n$, of $\MR{sp}(2n)$ and $\MR{o}(2n)$ types
in the near future.

\section*{Appendix}

\subsection*{A1\quad Commutation relations in the RTT--algebra $\MC{A}_n$}

The RTT--equation for the RTT--algebra $\MC{A}_n$ leads to
the commutation relations
$$
\begin{array}{l}
T^i_k(x)T^r_s(y)+g(x,y)T^r_k(x)T^i_s(y)=
T^r_s(y)T^i_k(x)+g(x,y)T^r_k(y)T^i_s(x)\\[4pt]
T^{-i}_{-k}(x)T^{-r}_{-s}(y)+g(x,y)T^{-r}_{-k}(x)T^{-i}_{-s}(y)=
T^{-r}_{-s}(y)T^{-i}_{-k}(x)+g(x,y)T^{-r}_{-k}(y)T^{-i}_{-s}(x)\\[4pt]
T^i_k(x)T^{-r}_{-s}(y)-\delta^{i,r}k(x,y){\tsum_{p=1}^n}T^p_k(x)T^{-p}_{-s}(y)=
T^{-r}_{-s}(y)T^i_k(x)-\delta_{k,s}k(x,y){\tsum_{p=1}^n}T^{-r}_{-p}(y)T^i_p(x)\\[4pt]
T^{-i}_{-k}(x)T^r_s(y)-\delta^{i,r}h(x,y){\tsum_{p=1}^n}T^{-p}_{-k}(x)T^p_s(y)=
T^r_s(y)T^{-i}_{-k}(x)-\delta_{k,s}h(x,y){\tsum_{p=1}^n}T^r_p(y)T^{-i}_{-p}(x)\\[4pt]
T^i_k(x)T^r_s(y)+g(y,x)T^i_s(x)T^r_k(y)=
T^r_s(y)T^i_k(x)+g(y,x)T^i_s(y)T^r_k(x)\\[4pt]
T^{-i}_{-k}(x)T^{-r}_{-s}(y)+g(y,x)T^{-i}_{-s}(x)T^{-r}_{-k}(y)=
T^{-r}_{-s}(y)T^{-i}_{-k}(x)+g(y,x)T^{-i}_{-s}(y)T^{-r}_{-k}(x)\\[4pt]
T^i_k(x)T^{-r}_{-s}(y)-\delta_{k,s}h(y,x){\tsum_{p=1}^n}T^i_p(x)T^{-r}_{-p}(y)=
T^{-r}_{-s}(y)T^i_k(x)-\delta^{i,r}h(y,x){\tsum_{p=1}^n}T^{-p}_{-s}(y)T^p_k(x)\\[4pt]
T^{-i}_{-k}(x)T^r_s(y)-\delta_{k,s}k(y,x){\tsum_{p=1}^n}T^{-i}_{-p}(x)T^r_p(y)=
T^r_s(y)T^{-i}_{-k}(x)-\delta^{i,r}k(y,x){\tsum_{p=1}^n}T^p_s(y)T^{-p}_{-k}(x)
\end{array}
$$

\subsection*{A2\quad Action of the operators $T^{\pm n}_{\pm n}(x)$ and
$\tilde{\MB{T}}^{(\pm)}(x)$ on the Bethe vectors}

First, we will rewrite the commutation relations using the operators action for
$P=Q=1$.

\smallskip

\underbar{\textbf{Lemma 1.}} In the RTT--algebra $\MC{A}_n$ the
following relations are true:
$$
\begin{array}{l}
T^n_n(x)\Bigl<\MB{b}^{(+)}_{1^*}(v),\MB{e}_k\Bigr>=
f(v,x)\Bigl<\MB{b}^{(+)}_{1^*}(v),\MB{e}_k\Bigr>T^n_n(x)-
g(v,x)\Bigl<\MB{b}^{(+)}_{1^*}(x),\MB{e}_k\Bigr>T^n_n(v)\\[6pt]
T^{-n}_{-n}(x)\Bigl<\MB{b}^{(-)}_1(w),\MB{f}^{-r}\Bigr>=
f(x,w)\Bigl<\MB{b}^{(-)}_1(w),\MB{f}^{-r}\Bigr>T^{-n}_{-n}(x)-
g(x,w)\Bigl<\MB{b}^{(-)}_1(x),\MB{f}^{-r}\Bigr>T^{-n}_{-n}(w)\\[6pt]
\tilde{\MB{T}}^{(+)}_0(x)\Bigl<\MB{b}^{(+)}_{1^*}(v),\MB{e}_k\Bigr>=
f(x,v)\Bigl<\MB{b}^{(+)}_{1^*}(v),\tilde{\MB{T}}^{(+)}_0(x)\wh{\MB{R}}^{(+,+)}_{0_+,1_+}(x,v)
\bigl(\MB{I}_{0_+}\otimes\MB{e}_k\bigr)\Bigr>-\\[4pt]
\hskip40mm-
g(x,v)\Bigl<\MB{b}^{(+)}_{1^*}(x),\tilde{\MB{T}}^{(+)}_0(v)\wh{\BB{R}}^{(+,+)}_{0_+,1_+}
\bigl(\MB{I}_{0_+}\otimes\MB{e}_k\bigr)\Bigr>\\[6pt]
\tilde{\MB{T}}^{(-)}_0(x)\Bigl<\MB{b}^{(-)}_1(w),\MB{f}^{-r}\Bigr>=
f(w,x)\Bigl<\MB{b}^{(-)}_1(w),\wh{\MB{R}}^{(-,-)}_{0_-,1^*_-}(x,w)\tilde{\MB{T}}^{(-)}_0(x)
\bigl(\tilde{\MB{I}}_-\otimes\MB{f}^{-r}\bigr)\Bigr>-\\[4pt]
\hskip40mm-
g(w,x)\Bigl<\MB{b}^{(-)}_1(x),\wh{\R}^{(-,-)}_{0_-,1^*_-}\tilde{\MB{T}}^{(-)}_0(w)
\bigl(\tilde{\MB{I}}_-\otimes\MB{f}^{-r}\bigr)\Bigr>\\[6pt]
T^n_n(x)\Bigl<\MB{b}^{(-)}_1(w),\MB{f}^{-r}\Bigr>=
\dfrac{\tilde{h}(w,x)}{h(w,x)}\Bigl<\MB{b}^{(-)}_1(w),\MB{f}^{-r}\Bigr>T^n_n(x)+\\[4pt]
\hskip40mm+ \tilde{h}(w,x)\Tr_{0}\Bigl<\MB{b}^{(+)}_{1^*}(x),
\wh{\BB{P}}^{(+,-)}_{1_+,1^*_-}\wh{\R}^{(-,-)}_{0_-,1^*_-}\tilde{\MB{T}}^{(-)}_0(w)
\bigl(\tilde{\MB{I}}_-\otimes\MB{f}^{-r}\bigr)\Bigr>\\[6pt]
T^{-n}_{-n}(x)\Bigl<\MB{b}^{(+)}_{1^*}(v),\MB{e}_k\Bigr>=
\dfrac{\tilde{h}(x,v)}{h(x,v)}\Bigl<\MB{b}^{(+)}_{1^*}(v),\MB{e}_k\Bigr>T^{-n}_{-n}(x)+\\[4pt]
\hskip40mm+ \tilde{h}(x,v)\Tr_{0}\Bigl<\MB{b}^{(-)}_1(x),
\wh{\BB{P}}^{(-,+)}_{1^*_-,1_+}\tilde{\MB{T}}^{(+)}_0(v)\wh{\BB{R}}^{(+,+)}_{0_+,1_+}
\bigl(\tilde{\MB{I}}_+\otimes\MB{e}_k\bigr)\Bigr>\\[6pt]
\tilde{\MB{T}}^{(+)}_0(x)\Bigl<\MB{b}^{(-)}_1(w),\MB{f}^{-r}\Bigr>=
\Bigl<\MB{b}^{(-)}_1(w),\wh{\MB{R}}^{(+,-)}_{0_+,1^*_-}(x,w)\tilde{\MB{T}}^{(+)}_0(x)
\bigl(\tilde{\MB{I}}_+\otimes\MB{f}^{-r}\bigr)\Bigr>-\\[4pt]
\hskip40mm-
\tilde{h}(w,x)\Bigl<\MB{b}^{(+)}_{1^*}(x),\wh{\BB{P}}^{(+,-)}_{1_+,1^*_-}\wh{\R}^{(+,-)}_{0_+,1^*_-}
\bigl(\tilde{\MB{I}}_+\otimes\MB{f}^{-r}\bigr)\Bigr>T^{-n}_{-n}(w)\\[6pt]
\tilde{\MB{T}}^{(-)}_0(x)\Bigl<\MB{b}^{(+)}_{1^*}(v),\MB{e}_k\Bigr>=
\Bigl<\MB{b}^{(+)}_{1^*}(v),\tilde{\MB{T}}^{(-)}_0(x)\wh{\MB{R}}^{(-,+)}_{0_-,1_+}(x,v)
\bigl(\tilde{\MB{I}}_-\otimes\MB{e}_k\bigr)\Bigr>-\\[4pt]
\hskip40mm-
\tilde{h}(x,v)\Bigl<\MB{b}^{(-)}_1(x),\wh{\BB{P}}^{(-,+)}_{1^*_-,1_+}\wh{\BB{R}}^{(-,+)}_{0_-1_+}
\bigl(\tilde{\MB{I}}_-\otimes\MB{e}_k\bigr)\Bigr>T^n_n(v)
\end{array}
$$
where
$$
\begin{array}{ll}
\wh{\R}^{(+,+)}_{0_+,1_+}= \tilde{\MB{R}}^{(+,+)}_{0_+,1_+}(x,x)=
{\tsum_{i,k=1}^{n-1}}\MB{E}^i_k\otimes\MB{E}^k_i\,,\qquad&
\wh{\R}^{(+,-)}_{0_+,1^*_-}={\tsum_{r,s=1}^{n-1}}\MB{E}^r_s\otimes\MB{F}^{-r}_{-s}\\[6pt]
\wh{\R}^{(-,-)}_{0_-,1^*_-}= \wh{\MB{R}}^{(-,-)}_{0_-,1^*_-}(w,w)=
{\tsum_{r,s=1}^{n-1}}\MB{E}^{-r}_{-s}\otimes\MB{F}^{-s}_{-r}\,,\qquad&
\wh{\R}^{(-,+)}_{0_-,1_+}={\tsum_{i,k=1}^{n-1}}\MB{E}^{-i}_{-k}\otimes\MB{E}^i_k
\end{array}
$$
and $\wh{\BB{P}}^{(+,-)}_{1_+,1^*_-}$,
$\wh{\BB{P}}^{(-,+)}_{1^*_-,1_+}$ are the linear mappings
$\wh{\BB{P}}^{(+,-)}_{1_+,1^*_-}:\MC{V}^*_{1^*_-}\to\MC{V}_{1_+}$,
$\wh{\BB{P}}^{(-,+)}_{1^*_-,1_+}:\MC{V}_{1_+}\to\MC{V}^*_{1^*_-}$
defined by
$$
\wh{\BB{P}}^{(+,-)}_{1_+,1^*_-}\MB{f}^{-r}=\MB{e}_r\,,\hskip15mm
\wh{\BB{P}}^{(-,+)}_{1^*_-,1_+}\MB{e}_k=\MB{f}^{-k}.
$$

\medskip

\textsc{Proof:} The first two equations are only otherwise written
commutation relations
$$
\begin{array}{l}
T^n_n(x)T^n_k(v)=f(v,x)T^n_k(v)T^n_n(x)-g(v,x)T^n_k(x)T^n_n(v)\,,\\[4pt]
T^{-n}_{-n}(x)T^{-r}_{-n}(w)=
f(x,w)T^{-r}_{-n}(w)T^{-n}_{-n}(x)-g(x,w)T^{-r}_{-n}(x)T^{-n}_{-n}(w)
\end{array}
$$
and the third and fourth relationships are the matrix notation of the commutation relations
$$
\begin{array}{l}
T^r_s(x)T^n_k(v)=
T^n_k(v)T^r_s(x)+g(x,v)T^n_s(v)T^r_k(x)-g(x,v)T^n_s(x)T^r_k(v)\,,\\[4pt]
T^{-i}_{-k}(x)T^{-r}_{-n}(w)=
T^{-r}_{-n}(w)T^{-i}_{-k}(x)+g(w,x)T^{-i}_{-n}(w)T^{-r}_{-k}(x)-g(w,x)T^{-i}_{-n}(x)T^{-r}_{-k}(w)\,.
\end{array}
$$

To prove the fifth relation, we first use the commutation relation
\begin{equation}
 \label{L4-1}
\begin{array}{l}
T^n_n(x)T^{-r}_{-n}(w)=T^{-r}_{-n}(w)T^n_n(x)-
k(x,w){\tsum_{p=1}^n}T^{-r}_{-p}(w)T^n_p(x)=\\[6pt]
\hskip20mm= \Bigl(1-k(x,w)\Bigr)T^{-r}_{-n}(w)T^n_n(x)-
k(x,w){\tsum_{p=1}^{n-1}}T^{-r}_{-p}(w)T^n_p(x)\,.
\end{array}
\end{equation}
If we sum the commutation relations
$$
T^{-r}_{-k}(w)T^n_k(x)= T^n_k(x)T^{-r}_{-k}(w)-
h(w,x){\tsum_{p=1}^n}T^n_p(x)T^{-r}_{-p}(w)
$$
over $k=1,\,\ldots,\,n-1$, we find that
$$
{\tsum_{p=1}^{n-1}}T^{-r}_{-p}(w)T^n_p(x)=
\Bigl(1-(n-1)h(w,x)\Bigr){\tsum_{p=1}^{n-1}}T^n_p(x)T^{-r}_{-p}(w)-
(n-1)h(w,x)T^n_n(x)T^{-r}_{-n}(w)\,
$$
When we substitute this equality into (\ref{L4-1}), we get
\begin{equation}
 \label{L4-2}
T^n_n(x)T^{-r}_{-n}(w)=
\Bigl(1+\tilde{h}(w,x)\Bigr)T^{-r}_{-n}(w)T^n_n(x)+
\tilde{h}(w,x){\tsum_{p=1}^{n-1}}T^n_p(x)T^{-r}_{-p}(w)
\end{equation}
which is another notation of the fifth relationship.

To prove the sixth relation, we use the commutation relations
\begin{equation}
 \label{L4-3}
T^{-n}_{-n}(x)T^n_k(v)= \Bigl(1-k(v,x)\Bigr)T^n_k(v)T^{-n}_{-n}(x)-
k(v,x){\tsum_{p=1}^{n-1}}T^p_k(v)T^{-p}_{-n}(x)\,.
\end{equation}
If we sum the commutation relations
$$
T^i_k(v)T^{-i}_{-n}(x)=
T^{-i}_{-n}(x)T^i_k(v)-h(x,v)T^{-n}_{-n}(x)T^n_k(v)-
h(x,v){\tsum_{p=1}^{n-1}}T^{-p}_{-n}(x)T^p_k(v)\,,
$$
over $i=1,\,\ldots,\,n-1$, we obtain
$$
{\tsum_{p=1}^{n-1}}T^p_k(v)T^{-p}_{-n}(x)=
\Bigl(1-(n-1)h(x,v)\Bigr){\tsum_{p=1}^{n-1}}T^{-p}_{-n}(x)T^p_k(v)-
(n-1)h(x,v)T^{-n}_{-n}(x)T^n_k(v)\,.
$$
When we substitute this relation into (\ref{L4-3}), we get
\begin{equation}
 \label{L4-4}
T^{-n}_{-n}(x)T^n_k(v)=
\Bigl(1+\tilde{h}(x,v)\Bigr)T^n_k(v)T^{-n}_{-n}(x)+
\tilde{h}(x,v){\tsum_{p=1}^{n-1}}T^{-p}_{-n}(x)T^p_k(v)
\end{equation}
which can be written in the form shown in Lemma.

To prove the seventh and eighth relationships,
we first use the commutation relations
$$
\begin{array}{l}
T^i_k(x)T^{-r}_{-n}(w)=
T^{-r}_{-n}(w)T^i_k(x)-\delta^{i,r}h(w,x)T^{-n}_{-n}(w)T^n_k(x)-
\delta^{i,r}h(w,x){\tsum_{p=1}^{n-1}}T^{-p}_{-n}(w)T^p_k(x)\\[6pt]
T^{-r}_{-s}(x)T^n_k(v)=
T^n_k(v)T^{-r}_{-s}(x)-\delta_{k,s}h(x,v)T^n_n(v)T^{-r}_{-n}(x)-
\delta_{k,s}h(x,v){\tsum_{p=1}^{n-1}}T^n_p(v)T^{-r}_{-p}(x)\,.
\end{array}
$$
Using equations (\ref{L4-4}) and (\ref{L4-2}),
we obtain
$$
\begin{array}{l}
T^i_k(x)T^{-r}_{-n}(w)= T^{-r}_{-n}(w)T^i_k(x)-
\delta^{i,r}\tilde{h}(w,x){\tsum_{p=1}^{n-1}}T^{-p}_{-n}(w)T^p_k(x)-
\delta^{i,r}\tilde{h}(w,x)T^n_k(x)T^{-n}_{-n}(w)\\[6pt]
T^{-r}_{-s}(x)T^n_k(v)= T^n_k(v)T^{-r}_{-s}(x)-
\delta_{k,s}\tilde{h}(x,v){\tsum_{p=1}^{n-1}}T^n_p(v)T^{-r}_{-p}(x)-
\delta_{k,s}\tilde{h}(x,v)T^{-r}_{-n}(x)T^n_n(v)
\end{array}
$$
which are other notations of the last two equations of Lemma.
\qed

\bigskip

Using Lemma 1, it is relatively easy to find members in which $x$
is exchanged with the first component of the vectors $\vec{v}$ and $\vec{w}$.
The members, in which $x$ is interchanged with other components of these vectors,
can be found by switching the corresponding component to the first place of vectors.

The following Lemma gives a suitable notation of the commutation relations that we use.

\smallskip

\underbar{\textbf{Lemma 2.}} In the RTT--algebra $\MC{A}_n$
the following relations apply:
$$
\begin{array}{l}
\Bigl<\MB{b}^{(+)}_{1^*}(x)\MB{b}^{(+)}_{2^*}(y),\MB{e}_i\otimes\MB{e}_k\Bigr>=
\Bigl<\MB{b}^{(+)}_{2^*}(y)\MB{b}^{(+)}_{1^*}(x),
\wh{\MB{R}}^{(+,+)}_{1_+,2_+}(x,y)\bigl(\MB{e}_i\otimes\MB{e}_k\bigr)\Bigr>\\[4pt]
\Bigl<\MB{b}^{(-)}_1(x)\MB{b}^{(-)}_2(y),\MB{f}^{-r}\otimes\MB{f}^{-s}\Bigr>=
\Bigl<\MB{b}^{(-)}_2(y)\MB{b}^{(-)}_1(x),\wh{\MB{R}}^{(-,-)}_{1^*_-,2^*_-}(y,x)
\bigl(\MB{f}^{-r}\otimes\MB{f}^{-s}\bigr)\Bigr>\\[4pt]
\Bigl<\MB{b}^{(+)}_{1^*}(x)\MB{b}^{(+)}_{2^*}(y),\MB{e}_i\otimes\MB{e}_k\Bigr>=
\Bigl<\MB{b}^{(+)}_{2^*}(x)\MB{b}^{(+)}_{1^*}(y),
\wh{\R}^{(+,+)}_{1_+,2_+}\bigl(\MB{e}_i\otimes\MB{e}_k\bigr)\Bigr>\\[4pt]
\Bigl<\MB{b}^{(-)}_1(x)\MB{b}^{(-)}_2(y),\MB{f}^{-r}\otimes\MB{f}^{-s}\Bigr>=
\Bigl<\MB{b}^{(-)}_2(x)\MB{b}^{(-)}_1(y),
\wh{\R}^{(-,-)}_{1^*_-,2^*_-}
\bigl(\MB{f}^{-r}\otimes\MB{f}^{-s}\bigr)\Bigr>\\[4pt]
\MB{b}^{(+)}_{1^*}(x)\MB{b}^{(-)}_2(y)=
\MB{b}^{(-)}_2(y)\MB{b}^{(+)}_{1^*}(x)\,,\\[4pt]
\Bigl<\MB{b}^{(+)}_{1^*}(x)\MB{b}^{(-)}_2(y),\MB{e}_k\otimes\MB{f}^{-r}\Bigr>=
\Bigl<\MB{b}^{(+)}_{2^*}(x)\MB{b}^{(-)}_1(y),
\wh{\BB{P}}^{(-,+)}_{1^*_-,1_+}\wh{\R}^{(+,+)}_{1_+,2_+}\wh{\BB{P}}^{(+,-)}_{2_+,2^*_-}
\bigl(\MB{e}_k\otimes\MB{f}^{-r}\bigr)\Bigr>=\\[4pt]
\hskip45mm= \Bigl<\MB{b}^{(+)}_{2^*}(x)\MB{b}^{(-)}_1(y),
\wh{\BB{P}}^{(+,-)}_{2_+,2^*_-}\wh{\R}^{(-,-)}_{1^*_-,2^*_-}
\wh{\BB{P}}^{(-,+)}_{1^*_-,1_+}\bigl(\MB{e}_k\otimes\MB{f}^{-r}\bigr)\Bigr>\\[4pt]
\Bigl<\MB{b}^{(+)}_{2^*}(x)\MB{b}^{(-)}_1(y),\MB{f}^{-r}\otimes\MB{e}_k\Bigr>=
\Bigl<\MB{b}^{(+)}_{1^*}(x)\MB{b}^{(-)}_2(y),
\wh{\BB{P}}^{(-,+)}_{2^*_-,2_+}\wh{\R}^{(+,+)}_{1_+,2_+}
\wh{\BB{P}}^{(+,-)}_{1_+,1^*_-}\bigl(\MB{f}^{-r}\otimes\MB{e}_k\bigr)\Bigr>=\\[4pt]
\hskip45mm= \Bigl<\MB{b}^{(+)}_{1^*}(x)\MB{b}^{(-)}_2(y),
\wh{\BB{P}}^{(+,-)}_{1_+,1^*_-}\wh{\R}^{(-,-)}_{1^*_-,2^*_-}
\wh{\BB{P}}^{(-,+)}_{2^*_-,2_+}\bigl(\MB{f}^{-r}\otimes\MB{e}_k\bigr)\Bigr>
\end{array}
$$
where
$$
\begin{array}{l}
\wh{\MB{R}}^{(-,-)}_{1^*_-,2^*_-}(y,x)=
\dfrac1{f(y,x)}\Bigl(\tilde{\MB{I}}^*_-\otimes\tilde{\MB{I}}^*_-+
g(y,x){\tsum_{r,s=1}^{n-1}}\MB{F}^{-r}_{-s}\otimes\MB{F}^{-s}_{-r}\Bigr)\\[6pt]
\wh{\R}^{(-,-)}_{1^*_-,2^*_-}=\wh{\MB{R}}^{(-,-)}_{1^*_-,2^*_-}(x,x)=
{\tsum_{r,s=1}^{n-1}}\MB{F}^{-r}_{-s}\otimes\MB{F}^{-s}_{-r}
\end{array}
$$

\textsc{Proof:}
The first and second relations are the transcripts of the commutation relations
$$
\begin{array}{l}
f(x,y)T^n_i(x)T^n_k(y)=T^n_k(y)T^n_i(x)+g(x,y)T^n_i(y)T^n_k(x)\\[4pt]
f(y,x)T^{-r}_{-n}(x)T^{-s}_{-n}(y)=
T^{-s}_{-n}(y)T^{-r}_{-n}(x)+g(y,x)T^{-r}_{-n}(y)T^{-s}_{-n}(x)\,,
\end{array}
$$
the fifth relation is an otherwise written commutation relation
$$
T^n_k(x)T^{-r}_{-n}(y)=T^{-r}_{-n}(y)T^n_k(x)
$$
and the other equations are the identities. \qed

\bigskip

To write the operators' action $T^{\pm n}_{\pm n}(x)$ and
$\tilde{\MB{T}}^{(\pm)}(x)$ on the Bethe vectors with the general
$\vec{v}$ and $\vec{w}$, we prefer to introduce
$$
\begin{array}{l}
\MB{b}^{(+)}_{k^*;1^*,\ldots,P^*}(x;\vec{v}_k)=
\MB{b}^{(+)}_{k^*}(x)\MB{b}^{(+)}_{1^*,\ldots,\wh{k}^*,\ldots,P^*}(\vec{v}_k)\\[4pt]
\MB{b}^{(-)}_{r;1,\ldots,Q}(x;\vec{w}_r)=
\MB{b}^{(-)}_r(x)\MB{b}^{(-)}_{1,\ldots,\wh{r},\ldots,Q}(\vec{w}_r)\\[4pt]
\MB{b}^{(+)}_{1^*,\ldots,\wh{k}^*,\ldots,P^*}(\vec{v}_k)=
\MB{b}^{(+)}_{1^*}(v_1)\ldots\MB{b}^{(+)}_{(k-1)^*}(v_{k-1})
\MB{b}^{(+)}_{(k+1)^*}(v_{k+1})\ldots\MB{b}^{(+)}_{P^*}(v_P)\\[4pt]
\MB{b}^{(-)}_{1,\ldots,\wh{r},\ldots,Q}(\vec{w}_r)=
\MB{b}^{(-)}_1(w_1)\ldots\MB{b}^{(-)}_{r-1}(w_{r-1})
\MB{b}^{(-)}_{r+1}(w_{r+1})\ldots \MB{b}^{(-)}_Q(w_Q)\\[4pt]
\wh{\MB{R}}^{(+,+)}_{1,\ldots,k}(\vec{v})=
\wh{\MB{R}}^{(+,+)}_{1_+,k_+}(v_1,v_k)
\wh{\MB{R}}^{(+,+)}_{2_+,k_+}(v_2,v_k)\ldots
\wh{\MB{R}}^{(+,+)}_{(k-1)_+,k_+}(v_{k-1},v_k)\\[4pt]
\wh{\MB{R}}^{(-,-)}_{1^*,\ldots,r^*}(\vec{w})=
\wh{\MB{R}}^{(-,-)}_{1^*_-,r^*_-}(w_r,w_1)
\wh{\MB{R}}^{(-,-)}_{2^*_-,r^*_-}(w_r,w_2)\ldots
\wh{\MB{R}}^{(-,-)}_{(r-1)^*_-,r^*_-}(w_r,w_{r-1})\\[4pt]
\wh{\MB{T}}^{(+,+)}_{0;1,\ldots,P}(x;\vec{v})=
\tilde{\MB{T}}^{(+)}_0(x)\wh{\MB{R}}^{(+,+)}_{0;1,\ldots,P}(x;\vec{v})\\[4pt]
\wh{\MB{T}}^{(-,-)}_{0;1^*,\ldots,Q^*}(x;\vec{w})=
\wh{\MB{R}}^{(-,-)}_{0;1^*,\ldots,Q^*}(x;\vec{w})\tilde{\MB{T}}^{(-)}_0(x)\\[4pt]
\wh{\BB{T}}^{(+,+)}_{k;0;1,\ldots,P}(\vec{v})=
\wh{\MB{T}}^{(+,+)}_{0;1,\ldots,P}(v_k;\vec{v})=
\tilde{\MB{T}}^{(+)}_0(v_k)\wh{\MB{R}}^{(+,+)}_{0;1,\ldots,P}(v_k;\vec{v})\\[4pt]
\wh{\BB{T}}^{(-,-)}_{r;0;1^*,\ldots,Q^*}(\vec{w})=
\wh{\MB{T}}^{(-,-)}_{0;1^*,\ldots,Q^*}(w_r;\vec{w})=
\wh{\MB{R}}^{(-,-)}_{0;1^*,\ldots,Q^*}(w_r;\vec{w})\tilde{\MB{T}}^{(-)}_0(w_r)\,.
\end{array}
$$

\underbar{\textbf{Lemma 3.}} For any $\vec{v}$ and $\vec{w}$
the relations
$$
\begin{array}{l}
T^n_n(x)\Bigl<\MB{b}^{(+)}_{1^*,\ldots,P^*}(\vec{v}),\MB{e}_{\vec{k}}\Bigr>=
F(\ol{v};x)\Bigl<\MB{b}^{(+)}_{1^*,\ldots,P^*}(\vec{v}),\MB{e}_{\vec{k}}\Bigr>T^n_n(x)-\\[4pt]
\hskip20mm- {\tsum_{v_k\in\ol{v}}}g(v_k,x)F(\ol{v}_k;v_k)
\Bigl<\MB{b}^{(+)}_{k^*;1^*,\ldots,P^*}(x;\vec{v}_k),
\wh{\MB{R}}^{(+,+)}_{1,\ldots,k}(\vec{v})\MB{e}_{\vec{k}}\Bigr>T^n_n(v_k)\\[6pt]
T^{-n}_{-n}(x)\Bigl<\MB{b}^{(-)}_{1,\ldots,Q}(\vec{w}),\MB{f}^{-\vec{r}}\Bigr>=
F(x;\ol{w})\Bigl<\MB{b}^{(-)}_{1,\ldots,Q}(\vec{w}),\MB{f}^{-\vec{r}}\Bigr>T^{-n}_{-n}(x)-\\[4pt]
\hskip20mm- {\tsum_{w_r\in\ol{w}}}g(x,w_r)F(w_r;\ol{w}_r)
\Bigl<\MB{b}^{(-)}_{r;1,\ldots,Q}(x;\vec{w}_r),
\wh{\MB{R}}^{(-,-)}_{1^*,\ldots,r^*}(\vec{w})\MB{f}^{-\vec{r}}\Bigr>T^{-n}_{-n}(w_r)\\[6pt]
\tilde{\MB{T}}^{(+)}_0(x)\Bigl<\MB{b}^{(+)}_{1^*,\ldots,P^*}(\vec{v}),\MB{e}_{\vec{k}}\Bigr>=
F(x;\ol{v})\Bigl<\MB{b}^{(+)}_{1^*,\ldots,P^*}(\vec{v}),
\wh{\MB{T}}^{(+,+)}_{0;1,\ldots,P}(x;\vec{v})\,\MB{e}_{\vec{k}}\Bigr>-\\[6pt]
\hskip20mm- {\tsum_{v_k\in\ol{v}}}g(x,v_k)F(v_k;\ol{v}_k)
\Bigl<\MB{b}^{(+)}_{k^*;1^*,\ldots,P^*}(x;\vec{v}_k),
\wh{\MB{R}}^{(+,+)}_{1,\ldots,k}(\vec{v})
\wh{\BB{T}}^{(+,+)}_{k;0;1,\ldots,P}(\vec{v})\,\MB{e}_{\vec{k}}\Bigr>\\[6pt]
\tilde{\MB{T}}^{(-)}_0(x)\Bigl<\MB{b}^{(-)}_{1,\ldots,Q}(\vec{w}),\MB{f}^{-\vec{r}}\Bigr>=
F(\ol{w};x)\Bigl<\MB{b}^{(-)}_{1,\ldots,Q}(\vec{w}),
\wh{\MB{T}}^{(-,-)}_{0;1^*,\ldots,Q^*}(x;\vec{w})\,\MB{f}^{-\vec{r}}\Bigr>-\\[6pt]
\hskip20mm- {\tsum_{w_r\in\ol{w}}}g(w_r,x)F(\ol{w}_r;w_r)
\Bigl<\MB{b}^{(-)}_{r;1,\ldots,Q}(x;\vec{w}_r),
\wh{\MB{R}}^{(-,-)}_{1^*,\ldots,r^*}(\vec{w})
\wh{\BB{T}}^{(-,-)}_{r;0;1^*,\ldots,Q^*}(\vec{w})\,\MB{f}^{-\vec{r}}\Bigr>
\end{array}
$$
hold in the RTT--algebra $\MC{A}_n$

\smallskip

\textsc{Proof:}
We can prove these statements by induction over the number of
elements $P$ and $Q$ of the sets $\ol{v}$ and $\ol{w}$.

For $P=Q=1$ these relations are proven in Lemma 1.

We will assume that the statement is valid for $P$ and $Q$ and denote
$\vec{v}=(v_1,\ldots,v_P,v_{P+1})$, $\vec{w}=(w_1,\ldots,w_Q,w_{Q+1})$,
$\vec{k}=(k_1,\ldots,k_P,k_{P+1})$ and $\vec{r}=(r_1,\ldots,r_Q,r_{Q+1})$.
According to the induction assumption and Lemma 1, we have
$$
\begin{array}{l}
T^n_n(x)\Bigl<\MB{b}^{(+)}_{1^*,\ldots,(P+1)^*}(\vec{v}),\MB{e}_{\vec{k}}\Bigr>=
T^n_n(x)\Bigl<\MB{b}^{(+)}_{1^*}(v_1),\MB{e}_{k_1}\Bigr>
\Bigl<\MB{b}^{(+)}_{2^*,\ldots,(P+1)^*}(\vec{v}_1),\MB{e}_{\vec{k}_1}\Bigr>=\\[6pt]
\hskip5mm= f(v_1,x)\Bigl<\MB{b}^{(+)}_{1^*}(v_1),\MB{e}_{k_1}\Bigr>
T^n_n(x)\Bigl<\MB{b}^{(+)}_{2^*,\ldots,(P+1)^*}(\vec{v}_1),\MB{e}_{\vec{k}_1}\Bigr>-\\[4pt]
\hskip20mm-
g(v_1,x)\Bigl<\MB{b}^{(+)}_{1^*}(x),\MB{e}_{k_1}\Bigr>
T^n_n(v_1)\Bigl<\MB{b}^{(+)}_{2^*,\ldots,(P+1)^*}(\vec{v}_1),\MB{e}_{\vec{k}_1}\Bigr>=\\[6pt]
\hskip5mm=
F(\ol{v};x)\Bigl<\MB{b}^{(+)}_{1^*,\ldots,(P+1)^*}(\vec{v}),\MB{e}_{\vec{k}}\Bigr>T^n_n(x)-\\[4pt]
\hskip20mm-
g(v_1,x)F(\ol{v}_1;v_1)
\Bigl<\MB{b}^{(+)}_{1^*;1^*,\ldots,(P+1)^*}(x,\vec{v}_1),\MB{e}_{\vec{k}}\Bigr>T^n_n(v_1)-\\[4pt]
\hskip10mm- {\tsum_{v_k\in\ol{v}_1}}F(\ol{v}_{1,k};v_k)
\biggl[g(v_k,x)f(v_1,x)
\Bigl<\MB{b}^{(+)}_{1^*}(v_1)\MB{b}^{(+)}_{k^*;2^*,\ldots,(P+1)^*}(x;\vec{v}_{1,k}),
\wh{\MB{R}}^{(+,+)}_{2,\ldots,k}(\vec{v}_1)\MB{e}_{\vec{k}}\Bigr>-\\[4pt]
\hskip30mm-
g(v_k,v_1)g(v_1,x)\Bigl<\MB{b}^{(+)}_{1^*}(x)\MB{b}^{(+)}_{k^*;2,\ldots,(P+1)^*}(v_1;\vec{v}_{1,k}),
\wh{\MB{R}}^{(+,+)}_{2,\ldots,k}(\vec{v}_1)\MB{e}_{\vec{k}}\Bigr>\biggr]T^n_n(v_k)\,,\\[9pt]
T^{-n}_{-n}(x)\Bigl<\MB{b}^{(-)}_{1,\ldots,Q}(\vec{w}),\MB{f}^{-\vec{r}}\Bigr>=
T^{-n}_{-n}(x)\Bigl<\MB{b}^{(-)}_1(w_1),\MB{f}^{-r_1}\Bigr>
\Bigl<\MB{b}^{(-)}_{2,\ldots,Q}(\vec{w}_1),\MB{f}^{-\vec{r}_1}\Bigr>=\\
\end{array}
$$

If we use in the first two equations the relations
$$
\begin{array}{l}
\Bigl<\MB{b}^{(+)}_{1^*}(v_1)\MB{b}^{(+)}_{k^*;2^*,\ldots,(P+1)^*}(x;\vec{v}_{1,k}),
\wh{\MB{R}}^{(+,+)}_{2,\ldots,k}(\vec{v}_1)\MB{e}_{\vec{k}}\Bigr>=\\[4pt]
\hskip30mm=
\Bigl<\MB{b}^{(+)}_{k^*;1^*,\ldots,(P+1)^*}(x;\vec{v}_k),
\wh{\MB{R}}^{(+,+)}_{1_+,k_+}(v_1,x)
\wh{\MB{R}}^{(+,+)}_{2,\ldots,k}(\vec{v}_1)\MB{e}_{\vec{k}}\Bigr>\\[6pt]
\Bigl<\MB{b}^{(+)}_{1^*}(x)\MB{b}^{(+)}_{k^*;2^*,\ldots,(P+1)^*}(v_1;\vec{v}_{1,k}),
\wh{\MB{R}}^{(+,+)}_{2,\ldots,k}(\vec{v}_1)\MB{e}_{\vec{k}}\Bigr>=\\[4pt]
\hskip30mm=
\Bigl<\MB{b}^{(+)}_{k^*;1^*,\ldots,(P+1)^*}(x;\vec{v}_k),
\wh{\R}^{(+,+)}_{1_+,k_+}\wh{\MB{R}}^{(+,+)}_{2,\ldots,k}(\vec{v}_1)\MB{e}_{\vec{k}}\Bigr>\\[6pt]
\Bigl<\MB{b}^{(-)}_1(w_1)\MB{b}^{(-)}_{r;2,\ldots,Q}(x;\vec{w}_{1,r}),
\wh{\MB{R}}^{(-,-)}_{2^*,\ldots,r^*}(\vec{w}_1)\MB{f}^{-\vec{r}}\Bigr>=\\[4pt]
\hskip30mm= \Bigl<\MB{b}^{(-)}_{r;1,\ldots,Q}(x;\vec{w}_r),
\wh{\MB{R}}^{(-,-)}_{1^*_-,r^*_-}(x,w_1)
\wh{\MB{R}}^{(-,-)}_{2^*,\ldots,r^*}(\vec{w}_1)\MB{f}^{-\vec{r}}\Bigr>\\[6pt]
\Bigl<\MB{b}^{(-)}_1(x)\MB{b}^{(-)}_{r;2,\ldots,Q}(w_1;\vec{w}_{1,r}),
\wh{\MB{R}}^{(-,-)}_{2^*,\ldots,r^*}(\vec{w}_1)\MB{f}^{-\vec{r}}\Bigr>=\\[4pt]
\hskip30mm= \Bigl<\MB{b}^{(-)}_{r;1,\ldots,Q}(x;\vec{w}_r),
\wh{\R}^{(-,-)}_{1^*_-,r^*_-}\wh{\MB{R}}^{(-,-)}_{2^*,\ldots,r^*}(\vec{w}_1)\MB{f}^{-\vec{r}}\Bigr>
\end{array}
$$
that result from Lemma 2, and compare the results with the first two
relations of the proven Lemma, we can see that it is enough to show
for any $v_k \in \ol{v}_1 $ and any $w_r \in \ol{w} _1$ the equalities
$$
\begin{array}{l}
g(v_k,x)f(v_1,v_k) \wh{\MB{R}}^{(+,+)}_{1_+,k_+}(v_1,v_k)=
g(v_k,x)f(v_1,x)\wh{\MB{R}}^{(+,+)}_{1_+,k_+}(v_1,x)-
g(v_k,v_1)g(v_1,x)\wh{\R}^{(+,+)}_{1_+,k_+}\,,\\[4pt]
g(x,w_r)f(w_r,w_1)
\wh{\MB{R}}^{(-,-)}_{1^*_-,r^*_-}(w_r,w_1)=\\[4pt]
\hskip20mm= g(x,w_r)f(x,w_1)
\wh{\MB{R}}^{(-,-)}_{1^*_-,r^*_-}(x,w_1)- g(x,w_1)g(w_1,w_r)
\wh{\R}^{(-,-)}_{1^*_-,r^*_-}\,.
\end{array}
$$
However, this is equivalent to the identities
$$
\begin{array}{l}
g(v_k,x)g(v_1,v_k)=g(v_k,x)g(v_1,x)-g(v_k,v_1)g(v_1,x)\,\\[4pt]
g(x,w_r)g(w_r,w_1)=g(x,w_r)g(x,w_1)-g(x,w_1)g(w_1,w_r)\,.
\end{array}
$$

To prove the third and fourth equality of Lemma, we use the relations
$$
\begin{array}{l}
\Bigl<\MB{b}^{(+)}_{1^*}(v_1)\MB{b}^{(+)}_{k^*;2^*,\ldots,(P+1)^*}(x;\vec{v}_{1,k}),
\wh{\MB{R}}^{(+,+)}_{2,\ldots,k}(\vec{v}_1)
\wh{\BB{T}}^{(+,+)}_{k;0;2,\ldots,P+1}(\vec{v}_1)
\wh{\MB{R}}^{(+,+)}_{0_+,1_+}(x,v_1)\,\MB{e}_{\vec{k}}\Bigr>=\\[4pt]
\hskip10mm=
\Bigl<\MB{b}^{(+)}_{k^*;1^*,\ldots,(P+1)^*}(x;\vec{v}_k),
\wh{\MB{R}}^{(+,+)}_{1,k}(v_1,x)\wh{\MB{R}}^{(+,+)}_{2,\ldots,k}(\vec{v}_1)
\wh{\BB{T}}^{(+,+)}_{k;0;2,\ldots,P+1}(\vec{v}_1)
\wh{\MB{R}}^{(+,+)}_{0_+,1_+}(x,v_1)\,\MB{e}_{\vec{k}}\Bigr>\\[6pt]
\Bigl<\MB{b}^{(+)}_{1^*}(x)\MB{b}^{(+)}_{k^*;2^*,\ldots,(P+1)^*}(v_1;\vec{v}_{1,k}),
\wh{\MB{R}}^{(+,+)}_{2,\ldots,k}(\vec{v}_1)
\wh{\BB{T}}^{(+,+)}_{k;0;2,\ldots,P+1}(\vec{v}_1)
\wh{\R}^{(+,+)}_{0_+,1_+}\,\MB{e}_{\vec{k}}\Bigr>=\\[4pt]
\hskip10mm=
\Bigl<\MB{b}^{(+)}_{k^*;1^*,\ldots,(P+1)^*}(x;\vec{v}_k),
\wh{\R}^{(+,+)}_{1,k}\wh{\MB{R}}^{(+,+)}_{2,\ldots,k}(\vec{v}_1)
\wh{\BB{T}}^{(+,+)}_{k;0;2,\ldots,P+1}(\vec{v}_1)
\wh{\R}^{(+,+)}_{0_+,1_+}\,\MB{e}_{\vec{k}}\Bigr>,\\[6pt]
\Bigl<\MB{b}^{(-)}_1(w_1)\MB{b}^{(-)}_{r;2,\ldots,Q+1}(x;\vec{w}_{1,r}),
\wh{\MB{R}}^{(-,-)}_{0_-,1^*_-}(x,w_1)\wh{\MB{R}}^{(-,-)}_{2^*,\ldots,r^*}(\vec{w}_1)
\wh{\BB{T}}^{(-,-)}_{r;0;2^*,\ldots,(Q+1)^*}(\vec{w}_1)\,\MB{f}^{-\vec{r}}\Bigr>=\\[4pt]
\hskip5mm= \Bigl<\MB{b}^{(-)}_{r;1,\ldots,Q+1}(x;\vec{w}_r),
\wh{\MB{R}}^{(-,-)}_{1^*_-,r^*_-}(x,w_1)
\wh{\MB{R}}^{(-,-)}_{0_-,1^*_-}(x,w_1)
\wh{\MB{R}}^{(-,-)}_{2^*,\ldots,r^*}(\vec{w}_1)
\wh{\BB{T}}^{(-,-)}_{r;0;2^*,\ldots,(Q+1)^*}(\vec{w}_1)\,\MB{f}^{-\vec{r}}\Bigr>\\[6pt]
\Bigl<\MB{b}^{(-)}_1(x)\MB{b}^{(-)}_{r;2,\ldots,Q+1}(w_1;\vec{w}_{1,r}),
\wh{\R}^{(-,-)}_{0_-,1^*_-}\wh{\MB{R}}^{(-,-)}_{2^*,\ldots,r^*}(\vec{w}_1)
\wh{\BB{T}}^{(-,-)}_{r;0;2^*,\ldots,(Q+1)^*}(\vec{w}_1)\,\MB{f}^{-\vec{r}}\Bigr>=\\
\end{array}
$$
that follow from Lemma 2. Then we get the equalities
$$
\begin{array}{l}
\tilde{\MB{T}}^{(+)}_0(x)\Bigl<\MB{b}^{(+)}_{1^*,\ldots,(P+1)^*}(\vec{v}),\MB{e}_{\vec{k}}\Bigr>=
F(x,\ol{v})\Bigl<\MB{b}^{(+)}_{1^*,\ldots,(P+1)^*}(\vec{v}_1),
\wh{\MB{T}}^{(+,+)}_{0;1,\ldots,P+1}(x;\vec{v}_1)\,\MB{e}_{\vec{k}}\Bigr>-\\[6pt]
\hskip15mm-
g(x,v_1)F(v_1,\ol{v}_1)
\Bigl<\MB{b}^{(+)}_{1^*;1^*,\ldots,(P+1)^*}(x;\vec{v}_1),
\wh{\BB{T}}^{(+,+)}_{1;0;1,\ldots,P+1}(\vec{v})\,\MB{e}_{\vec{k}}\Bigr>-\\[6pt]
\hskip15mm-
{\tsum_{v_k\in\ol{v}_1}}F(v_k,\ol{v}_{1,k})\biggl(g(x,v_k)f(x,v_1)
\Bigl<\MB{b}^{(+)}_{k^*;1^*,\ldots,(P+1)^*}(x;\vec{v}_k),\\[4pt]
\hskip50mm
\wh{\MB{R}}^{(+,+)}_{1,k}(v_1,x)\wh{\MB{R}}^{(+,+)}_{2,\ldots,k}(\vec{v}_1)
\wh{\BB{T}}^{(+,+)}_{k;0;2,\ldots,P+1}(\vec{v}_1)
\wh{\MB{R}}^{(+,+)}_{0_+,1_+}(x,v_1)\,\MB{e}_{\vec{k}}\Bigr>-\\[6pt]
\hskip35mm-
g(x,v_1)g(v_1,v_k)
\Bigl<\MB{b}^{(+)}_{k^*;1^*,\ldots,(P+1)^*}(x;\vec{v}_k),\\[4pt]
\hskip50mm
\wh{\R}^{(+,+)}_{1,k}\wh{\MB{R}}^{(+,+)}_{2,\ldots,k}(\vec{v}_1)
\wh{\BB{T}}^{(+,+)}_{k;0;2,\ldots,P+1}(\vec{v}_1)
\wh{\R}^{(+,+)}_{0_+,1_+}\,\MB{e}_{\vec{k}}\Bigr>\biggr)\\[9pt]
\tilde{\MB{T}}^{(-)}_0(x)\Bigl<\MB{b}^{(-)}_{1,\ldots,Q+1}(\vec{w}),\MB{f}^{-\vec{r}}\Bigr>=
F(\ol{w};x)\Bigl<\MB{b}^{(-)}_{1,\ldots,Q+1}(\vec{w}),
\wh{\MB{T}}^{(-,-)}_{0;1^*,\ldots,(Q+1)^*}(x;\vec{w})\,
\MB{f}^{-\vec{r}}\Bigr>-\\[6pt]
\hskip10mm-
g(w_1,x)F(\ol{w}_1,w_1)\Bigl<\MB{b}^{(-)}_{1;1,\ldots,Q+1}(x;\vec{w}_1),
\wh{\BB{T}}^{(-,-)}_{1;0;1^*,\ldots,(Q+1)^*}(\vec{w})\,\MB{f}^{-\vec{r}}\Bigr>-\\
\end{array}
$$
If we show that the relations
$$
\begin{array}{l}
g(x,v_k)f(v_k,v_1)\wh{\MB{R}}^{(+,+)}_{1,\ldots,k}(\vec{v})
\wh{\BB{T}}^{(+,+)}_{k;0;1,\ldots,P+1}(\vec{v})=\\[4pt]
\hskip10mm=
g(x,v_k)f(x,v_1)\wh{\MB{R}}^{(+,+)}_{1,k}(v_1,x)\wh{\MB{R}}^{(+,+)}_{2,\ldots,k}(\vec{v}_1)
\wh{\BB{T}}^{(+,+)}_{k;0;2,\ldots,P+1}(\vec{v}_1)
\wh{\MB{R}}^{(+,+)}_{0_+,1_+}(x,v_1)-\\[4pt]
\hskip35mm-
g(x,v_1)g(v_1,v_k)\wh{\R}^{(+,+)}_{1,k}\wh{\MB{R}}^{(+,+)}_{2,\ldots,k}(\vec{v}_1)
\wh{\BB{T}}^{(+,+)}_{k;0;2,\ldots,P+1}(\vec{v}_1)
\wh{\R}^{(+,+)}_{0_+,1_+}\,,\\[6pt]
g(w_r,x)f(w_1,w_r)\wh{\MB{R}}^{(-,-)}_{1^*,\ldots,r^*}(\vec{w})
\wh{\BB{T}}^{(-,-)}_{r;0;1^*,\ldots,(Q+1)^*}(\vec{w})=\\[4pt]
\hskip10mm=
g(w_r,x)f(w_1,x)\wh{\MB{R}}^{(-,-)}_{1^*_-,r^*_-}(x,w_1)
\wh{\MB{R}}^{(-,-)}_{0_-,1^*_-}(x,w_1)
\wh{\MB{R}}^{(-,-)}_{2^*,\ldots,r^*}(\vec{w}_1)
\wh{\BB{T}}^{(-,-)}_{r;0;2^*,\ldots,(Q+1)^*}(\vec{w}_1)-\\[4pt]
\hskip20mm-
g(w_r,w_1)g(w_1,x)\wh{\R}^{(-,-)}_{1^*_-,r^*_-}\wh{\R}^{(-,-)}_{0_-,1^*_-}
\wh{\MB{R}}^{(-,-)}_{2^*,\ldots,r^*}(\vec{w}_1)
\wh{\BB{T}}^{(-,-)}_{r;0;2^*,\ldots,(Q+1)^*}(\vec{w}_1),
\end{array}
$$
are valid for any $v_k\in\ol{v}_1$ and
$w_r\in\ol{w}_1$, the third and fourth equality in Lemma will hold.

Since, by definition
$$
\begin{array}{l}
\wh{\BB{T}}^{(+,+)}_{k;0;1,\ldots,P+1}(\vec{v})=
\tilde{\MB{T}}^{(+)}_0(v_k)\wh{\MB{R}}^{(+,+)}_{0;1,\ldots,P+1}(v_k;\vec{v})\\[4pt]
\wh{\BB{T}}^{(+,+)}_{k;0;2,\ldots,P+1}(\vec{v}_1)=
\tilde{\MB{T}}^{(+)}_0(v_k)\wh{\MB{R}}^{(+,+)}_{0;2,\ldots,P+1}(v_k;\vec{v}_1)\\[4pt]
\wh{\BB{T}}^{(-,-)}_{r;0;1^*,\ldots,(Q+1)^*}(\vec{w})=
\wh{\MB{R}}^{(-,-)}_{0;1^*,\ldots,(Q+1)^*}(w_r;\vec{w})\tilde{\MB{T}}^{(-)}_0(w_r)\\[4pt]
\wh{\BB{T}}^{(-,-)}_{r;0;2^*,\ldots,(Q+1)^*}(\vec{w}_1)=
\wh{\MB{R}}^{(-,-)}_{0;2^*,\ldots,(Q+1)^*}(w_r;\vec{w}_1)\tilde{\MB{T}}^{(-)}_0(w_r)\,
\end{array}
$$
it suffices to show that
$$
\begin{array}{l}
g(x,v_k)f(v_k,v_1) \wh{\MB{R}}^{(+,+)}_{1,\ldots,k}(\vec{v})
\wh{\MB{R}}^{(+,+)}_{0;1,\ldots,P+1}(v_k;\vec{v})=\\[4pt]
\hskip10mm= g(x,v_k)f(x,v_1)
\wh{\MB{R}}^{(+,+)}_{1,k}(v_1,x)\wh{\MB{R}}^{(+,+)}_{2,\ldots,k}(\vec{v}_1)
\wh{\MB{R}}^{(+,+)}_{0;2,\ldots,P+1}(v_k;\vec{v}_1)
\wh{\MB{R}}^{(+,+)}_{0_+,1_+}(x,v_1)-\\[4pt]
\hskip35mm- g(x,v_1)g(v_1,v_k)
\wh{\R}^{(+,+)}_{1,k}\wh{\MB{R}}^{(+,+)}_{2,\ldots,k}(\vec{v}_1)
\wh{\MB{R}}^{(+,+)}_{0;2,\ldots,P+1}(v_k;\vec{v}_1)
\wh{\R}^{(+,+)}_{0_+,1_+}\\[6pt]
g(w_r,x)f(w_1,w_r)\wh{\MB{R}}^{(-,-)}_{1^*,\ldots,r^*}(\vec{w})
\wh{\MB{R}}^{(-,-)}_{0;1^*,\ldots,(Q+1)^*}(w_r;\vec{w})=\\[4pt]
\hskip10mm=
g(w_r,x)f(w_1,x)\wh{\MB{R}}^{(-,-)}_{1^*_-,r^*_-}(x,w_1)
\wh{\MB{R}}^{(-,-)}_{0_-,1^*_-}(x,w_1)
\wh{\MB{R}}^{(-,-)}_{2^*,\ldots,r^*}(\vec{w}_1)
\wh{\MB{R}}^{(-,-)}_{0;2^*,\ldots,(Q+1)^*}(w_r;\vec{w}_1)-\\[4pt]
\hskip20mm-
g(w_r,w_1)g(w_1,x)\wh{\R}^{(-,-)}_{1^*_-,r^*_-}\wh{\R}^{(-,-)}_{0_-,1^*_-}
\wh{\MB{R}}^{(-,-)}_{2^*,\ldots,r^*}(\vec{w}_1)
\wh{\MB{R}}^{(-,-)}_{0;2^*,\ldots,(Q+1)^*}(w_r;\vec{w}_1)
\end{array}
$$
are true.

If we use the definitions of the products of R--matrices, we find that
it is enough to show
$$
\begin{array}{l}
g(x,v_k)f(v_k,v_1)
\wh{\MB{R}}^{(+,+)}_{1,k}(v_1,v_k)\ldots\wh{\MB{R}}^{(+,+)}_{k-1,k}(v_{k-1},v_k)\\[4pt]
\hskip60mm \wh{\R}^{(+,+)}_{0,k}
\wh{\MB{R}}^{(+,+)}_{0,k-1}(v_k,v_{k-1})\ldots
\wh{\MB{R}}^{(+,+)}_{0,1}(v_k,v_1)=\\[6pt]
\hskip10mm= g(x,v_k)f(x,v_1) \wh{\MB{R}}^{(+,+)}_{1,k}(v_1,x)
\wh{\MB{R}}^{(+,+)}_{2,k}(v_2,v_k)\ldots\wh{\MB{R}}^{(+,+)}_{k-1,k}(v_{k-1},v_k)\\[4pt]
\hskip60mm \wh{\R}^{(+,+)}_{0,k}
\wh{\MB{R}}^{(+,+)}_{0,k-1}(v_k,v_{k-1})\ldots
\wh{\MB{R}}^{(+,+)}_{0,2}(v_k,v_2)
\wh{\MB{R}}^{(+,+)}_{0,1}(x,v_1)-\\[6pt]
\hskip20mm- g(x,v_1)g(v_1,v_k) \wh{\R}^{(+,+)}_{1,k}
\wh{\MB{R}}^{(+,+)}_{2,k}(v_2,v_k)\ldots\wh{\MB{R}}^{(+,+)}_{k-1,k}(v_{k-1},v_k)\\[4pt]
\hskip60mm \wh{\R}^{(+,+)}_{0,k}
\wh{\MB{R}}^{(+,+)}_{0,k-1}(v_k,v_{k-1})\ldots
\wh{\MB{R}}^{(+,+)}_{0,2}(v_k,v_2) \wh{\R}^{(+,+)}_{0,1},\\[6pt]
g(w_r,x)f(w_1,w_r) \wh{\MB{R}}^{(-,-)}_{1^*_-,r^*_-}(w_r,w_1)\ldots
\wh{\MB{R}}^{(-,-)}_{(r-1)^*_-,r^*_-}(w_r,w_{r-1})\\[4pt]
\hskip40mm
\wh{\MB{R}}^{(-,-)}_{0_-,1^*_-}(w_r,w_1)\ldots\wh{\MB{R}}^{(-,-)}_{0_-,(r-1)^*_-}(w_r,w_{r-1})
\wh{\R}^{(-,-)}_{0_-,r^*_-}=\\[4pt]
\hskip10mm= g(w_r,x)f(w_1,x)
\wh{\MB{R}}^{(-,-)}_{1^*_-,r^*_-}(x,w_1)
\wh{\MB{R}}^{(-,-)}_{2^*_-,r^*_-}(w_r,w_2)\ldots
\wh{\MB{R}}^{(-,-)}_{(r-1)^*_-,r^*_-}(w_r,w_{r-1})\\[4pt]
\hskip40mm
\wh{\MB{R}}^{(-,-)}_{0_-,1^*_-}(x,w_1)\wh{\MB{R}}^{(-,-)}_{0_-,2^*_-}(w_r,w_2)\ldots
\wh{\MB{R}}^{(-,-)}_{0_-,(r-1)^*_-}(w_r,w_{r-1})
\wh{\R}^{(-,-)}_{0_-,r^*_-}-\\[4pt]
\hskip20mm- g(w_r,w_1)g(w_1,x)
\wh{\R}^{(-,-)}_{1^*_-,r^*_-}\wh{\MB{R}}^{(-,-)}_{2^*_-,r^*_-}(w_r,w_2)\ldots
\wh{\MB{R}}^{(-,-)}_{(r-1)^*_-,r^*_-}(w_r,w_{r-1})\\[4pt]
\hskip40mm
\wh{\R}^{(-,-)}_{0_-,1^*_-}\wh{\MB{R}}^{(-,-)}_{0_-,2^*_-}(w_r,w_2)\ldots
\wh{\MB{R}}^{(-,-)}_{0_-,(r-1)^*_-}(w_r,w_{r-1})\wh{\R}^{(-,-)}_{0_-,r^*_-}\,.
\end{array}
$$

When we use the Yang--Baxter equations
$$
\begin{array}{l}
\wh{\MB{R}}^{(+,+)}_{r_+,k_+}(v_r,v_k)\wh{\R}^{(+)}_{0_+,k_+}\wh{\MB{R}}^{(+,+)}_{0_+,r_+}(v_k,v_r)=
\wh{\MB{R}}^{(+,+)}_{0_+,r_+}(v_k,v_r)\wh{\R}^{(+)}_{0_+,k_+}\wh{\MB{R}}^{(+,+)}_{r_+,k_+}(v_r,v_k)\\[4pt]
\wh{\MB{R}}^{(-,-)}_{s^*_-,r^*_-}(w_r,w_s)\wh{\MB{R}}^{(-,-)}_{0_-,s^*_-}(w_r,w_s)\wh{\R}^{(-,-)}_{0_-,r^*_-}=
\wh{\R}^{(-,-)}_{0_-,r^*_-}\wh{\MB{R}}^{(-,-)}_{0_-,s^*_-}(w_r,w_s)\wh{\MB{R}}^{(-,-)}_{s^*_-,r^*_-}(w_r,w_s)
\end{array}
$$
several times, it can be found that it is enough to prove the relations
$$
\begin{array}{l}
g(x,v_k)f(v_k,v_1)
\wh{\MB{R}}^{(+,+)}_{1_+,k_+}(v_1,v_k)\wh{\R}^{(+,+)}_{0_+,k_+}\wh{\MB{R}}^{(+,+)}_{0_+,1_+}(v_k,v_1)=\\[4pt]
\hskip20mm= g(x,v_k)f(x,v_1)
\wh{\MB{R}}^{(+,+)}_{1_+,k_+}(v_1,x)\wh{\R}^{(+,+)}_{0_+,k_+}\wh{\MB{R}}^{(+,+)}_{0_+,1_+}(x,v_1)-\\[4pt]
\hskip40mm-
g(x,v_1)g(v_1,v_k)\wh{\R}^{(+,+)}_{1_+,k_+}\wh{\R}^{(+,+)}_{0_+,k_+}\wh{\R}^{(+,+)}_{0_+,1_+}\,,\\[6pt]
g(w_r,x)f(w_1,w_r)
\wh{\MB{R}}^{(-,-)}_{1^*_-,r^*_-}(w_r,w_1)\wh{\MB{R}}^{(-,-)}_{0_-,1^*_-}(w_r,w_1)\wh{\R}^{(-,-)}_{0_-,r^*_-}=\\[4pt]
\hskip20mm= g(w_r,x)f(w_1,x)
\wh{\MB{R}}^{(-,-)}_{1^*_-,r^*_-}(x,w_1)\wh{\MB{R}}^{(-,-)}_{0_-,1^*_-}(x,w_1)\wh{\R}^{(-,-)}_{0_-,r^*_-}-\\[4pt]
\hskip40mm- g(w_r,w_1)g(w_1,x)
\wh{\R}^{(-,-)}_{1^*_-,r^*_-}\wh{\R}^{(-,-)}_{0_-,1^*_-}\wh{\R}^{(-,-)}_{0_-,r^*_-},
\end{array}
$$
which can be verified by direct calculation.
\qed

\bigskip

\underbar{\textbf{Lemma 4.}} For any $\vec{v}$, $\vec{w}$ and
$\vec{k}$, $\vec{r}$ the following relations are true:
$$
\begin{array}{l}
T^n_n(x)\Bigl<\MB{b}^{(-)}_{1,\ldots,Q}(\vec{w}),\MB{f}^{-\vec{r}}\Bigr>=
F(\ol{w};x-n+1+\eta)
\Bigl<\MB{b}^{(-)}_{1,\ldots,Q}(\vec{w}),\MB{f}^{-\vec{r}}\Bigr>T^n_n(x)+\\[9pt]
\hskip20mm+ {\tsum_{w_s\in\ol{w}}}\tilde{h}(w_s,x)F(\ol{w}_s;w_s)
\Tr_{0}\biggl(\Bigl<\MB{b}^{(+)}_{s^*}(x)
\MB{b}^{(-)}_{1,\ldots\wh{s},\ldots,Q}(\vec{w}_s),\\[4pt]
\hskip60mm \wh{\BB{P}}^{(+,-)}_{s_+,s^*_-}
\wh{\MB{R}}^{(-,-)}_{1^*,\ldots,s^*}(\vec{w})
\wh{\BB{T}}^{(-,-)}_{s;0;1^*,\ldots,Q^*}(\vec{w})\MB{f}^{-\vec{r}}\Bigr>\biggr)\\[12pt]
T^{-n}_{-n}(x)\Bigl<\MB{b}^{(+)}_{1^*,\ldots,P^*}(\vec{v}),\MB{e}_{\vec{k}}\Bigr>=
F(x+n-1-\eta;\ol{v})
\Bigl<\MB{b}^{(+)}_{1^*,\ldots,P^*}(\vec{v}),\MB{e}_{\vec{k}}\Bigr>T^{-n}_{-n}(x)+\\[9pt]
\hskip20mm+
{\tsum_{v_{\ell}\in\ol{v}}}\tilde{h}(x,v_{\ell})F(v_{\ell};\ol{v}_{\ell})
\Tr_{0}\biggl(\Bigl<\MB{b}^{(+)}_{1^*,\ldots,\wh{\ell}^*,\ldots,P^*}(\vec{v}_{\ell})
\MB{b}^{(-)}_{\ell}(x),\\[4pt]
\hskip60mm \wh{\BB{P}}^{(-,+)}_{\ell^*_-,\ell_+}
\wh{\MB{R}}^{(+,+)}_{1,\ldots,\ell}(\vec{v})
\wh{\BB{T}}^{(+,+)}_{\ell;0;1,\ldots,P}(\vec{v})\MB{e}_{\vec{k}}\Bigr>\biggr)\\[12pt]
\tilde{\MB{T}}^{(+)}_0(x)\Bigl<\MB{b}^{(-)}_{1,\ldots,Q}(\vec{w}),\MB{f}^{-\vec{r}}\Bigr>=
\Bigl<\MB{b}^{(-)}_{1,\ldots,Q}(\vec{w}),
\wh{\MB{T}}^{(+,-)}_{0;1^*,\ldots,Q^*}(x;\vec{w})\MB{f}^{-\vec{r}}\Bigr>-\\[9pt]
\hskip20mm- {\tsum_{w_s\in\ol{w}}}\tilde{h}(w_s,x)F(w_s;\ol{w}_s)
\Bigl<\MB{b}^{(+)}_{s^*}(x)\MB{b}^{(-)}_{1,\ldots,\wh{s},\ldots,Q}(\vec{w}_s),\\
\end{array}
$$
where
$$
\wh{\MB{T}}^{(+,-)}_{0;1^*,\ldots,Q^*}(x;\vec{w})=
\wh{\MB{R}}^{(+,-)}_{0;1^*,\ldots,Q^*}(x;\vec{w})\tilde{\MB{T}}^{(+)}_0(x)\,,\qquad
\wh{\MB{T}}^{(-,+)}_{0;1,\ldots,P}(x;\vec{v})=
\tilde{\MB{T}}^{(-)}_0(x)
\wh{\MB{R}}^{(-,+)}_{0;1,\ldots,P}(x;\vec{v})\,.
$$

\smallskip

\textsc{Proof:} These statements can be proven by induction
according to the number of elements $P$ and $Q$ of the sets $\ol{v}$ and
$\ol{w}$. For $P=1$ and  $Q=1$, these statements are proved in Lemma
3.

Assume that these statements hold for $P$ and $Q$ and denote
$\vec{v}=(v_1,\ldots,v_{P+1})$, $\vec{w}=(w_1,\ldots,w_{Q+1})$,
$\vec{k}=(k_1,\ldots,k_{P+1})$ and $\vec{r}=(r_1,\ldots,r_{Q+1})$.

To show the first statement, we use the equality
$$
\begin{array}{l}
T^n_n(x)\Bigl<\MB{b}^{(-)}_{1,\ldots,Q+1}(\vec{w}),\MB{f}^{-\vec{r}}\Bigr>=
T^n_n(x)\Bigl<\MB{b}^{(-)}_1(w_1),\MB{f}^{-r_1}\Bigr>
\Bigl<\MB{b}^{(-)}_{2,\ldots,Q+1}(\vec{w}_1),\MB{f}^{-\vec{r}_1}\Bigr>=\\[6pt]
\hskip10mm=
\dfrac{\tilde{h}(w_1,x)}{h(w_1,x)}\Bigl<\MB{b}^{(-)}_1(w_1),\MB{f}^{-r_1}\Bigr>
T^n_n(x)\Bigl<\MB{b}^{(-)}_{2,\ldots,Q+1}(\vec{w}_1),\MB{f}^{-\vec{r}_1}\Bigr>+\\[4pt]
\hskip20mm+
\tilde{h}(w_1,x)\Tr_{0}\biggl(\Bigl<\MB{b}^{(+)}_{1^*}(x),
\wh{\BB{P}}^{(+,-)}_{1_+,1^*_-}
\wh{\R}^{(-,-)}_{0_-,1^*_-}\MB{f}^{-r_1}\Bigr>\tilde{\MB{T}}^{(-)}_0(w_1)
\Bigl<\MB{b}^{(-)}_{2,\ldots,Q+1}(\vec{w}_1),\MB{f}^{-\vec{r}_1}\Bigr>\biggr),
\end{array}
$$
which results from Lemma 1. Using the induction assumption and Lemma 3,
we will get
$$
\begin{array}{l}
T^n_n(x)\Bigl<\MB{b}^{(-)}_{1,\ldots,Q+1}(\vec{w}),\MB{f}^{-\vec{r}}\Bigr>=
F(\ol{w};x-n+1+\eta)
\Bigl<\MB{b}^{(-)}_{1,\ldots,Q+1}(\vec{w}),\MB{f}^{-\vec{r}}\Bigr>T^n_n(x)+\\[6pt]
\hskip20mm+
\tilde{h}(w_1,x)F(\ol{w}_1;w_1)\Tr_{0}
\biggl(\Bigl<\MB{b}^{(+)}_{1^*}(x)\MB{b}^{(-)}_{2,\ldots,Q+1}(\vec{w}_1),\\[4pt]
\hskip60mm
\wh{\BB{P}}^{(+,-)}_{1_+,1^*_-}\wh{\R}^{(-,-)}_{0_-,1^*_-}
\wh{\BB{T}}^{(-,-)}_{1;0;1^*,\ldots,(Q+1)^*}(\vec{w})\,\MB{f}^{-\vec{r}}\Bigr>\biggr)+\\[6pt]
\hskip20mm+
{\tsum_{w_s\in\ol{w}_1}}\dfrac{\tilde{h}(w_1,x)}{h(w_1,x)}\tilde{h}(w_s,x)F(\ol{w}_{1,s};w_s)
\Tr_{0}\biggl(\Bigl<\MB{b}^{(+)}_{s^*}(x)\MB{b}^{(-)}_{1,\ldots\wh{s},\ldots,Q+1}(\vec{w}_s),\\[4pt]
\hskip60mm
\wh{\BB{P}}^{(+,-)}_{s_+,s^*_-}\wh{\MB{R}}^{(-,-)}_{2^*,\ldots,s^*}(\vec{w}_1)
\wh{\BB{T}}^{(-,-)}_{s;0;2^*,\ldots,(Q+1)^*}(\vec{w}_1)\MB{f}^{-\vec{r}}\Bigr>\biggr)-\\[6pt]
\hskip20mm-
{\tsum_{w_s\in\ol{w}_1}}\tilde{h}(w_1,x)g(w_s,w_1)F(\ol{w}_{1,s};w_s)
\Tr_{0}\biggl(
\Bigl<\MB{b}^{(+)}_{1^*}(x)\MB{b}^{(-)}_{s;2,\ldots,Q+1}(w_1;\vec{w}_{1,s}),\\[4pt]
\hskip60mm
\wh{\BB{P}}^{(+,-)}_{1_+,1^*_-}\wh{\R}^{(-,-)}_{0_-,1^*_-}
\wh{\MB{R}}^{(-,-)}_{2^*,\ldots,s^*}(\vec{w}_1)
\wh{\BB{T}}^{(-,-)}_{s;0;2^*,\ldots,(Q+1)^*}(\vec{w}_1)\,\MB{f}^{-\vec{r}}\Bigr>\biggr).
\end{array}
$$
When we use Lemma 2 and the relationship
$\wh{\BB{P}}^{(-,+)}_{1^*_-,1_+}\wh{\BB{P}}^{(+,-)}_{1_+,1^*_-}=\tilde{\MB{I}}_{1^*_-}$,
we obtain the relation
$$
\begin{array}{l}
\Bigl<\MB{b}^{(+)}_{1^*}(x)\MB{b}^{(-)}_{s;2,\ldots,Q+1}(w_1;\vec{w}_{1,s}),
\wh{\BB{P}}^{(+,-)}_{1_+,1^*_-}\wh{\R}^{(-,-)}_{0_-,1^*_-}
\wh{\MB{R}}^{(-,-)}_{2^*,\ldots,s^*}(\vec{w}_1)
\wh{\BB{T}}^{(-,-)}_{s;0;2^*,\ldots,(Q+1)^*}(\vec{w}_1)\,\MB{f}^{-\vec{r}}\Bigr>=\\[4pt]
\hskip5mm=
\Bigl<\MB{b}^{(+)}_{s^*}(x)\MB{b}^{(-)}_{1,\ldots\wh{s},\ldots,Q+1}(\vec{w}_s),
\BB{P}^{(+,-)}_{s_+,s^*_-}\wh{\R}^{(-,-)}_{1^*_-,s^*_-}
\wh{\R}^{(-,-)}_{0_-,1^*_-}\wh{\MB{R}}^{(-,-)}_{2^*,\ldots,s^*}(\vec{w}_1)
\wh{\BB{T}}^{(-,-)}_{s;0;2^*,\ldots,(Q+1)^*}(\vec{w}_1)\,\MB{f}^{-\vec{r}}\Bigr>.
\end{array}
$$
So it is enough to show that for any $s=2,\ldots,Q+1$ we have
$$
\begin{array}{l}
\tilde{h}(w_s,x)f(w_1,w_s)
\wh{\MB{R}}^{(-,-)}_{1^*,\ldots,s^*}(\vec{w})
\wh{\BB{T}}^{(-,-)}_{s;0;1^*,\ldots,(Q+1)^*}(\vec{w})=\\[9pt]
\hskip10mm= \dfrac{\tilde{h}(w_1,x)}{h(w_1,x)}\tilde{h}(w_s,x)
\wh{\MB{R}}^{(-,-)}_{2^*,\ldots,s^*}(\vec{w}_1)
\wh{\BB{T}}^{(-,-)}_{s;0;2^*,\ldots,(Q+1)^*}(\vec{w}_1)-\\[9pt]
\hskip20mm- \tilde{h}(w_1,x)g(w_s,w_1) \wh{\R}^{(-,-)}_{1^*_-,s^*_-}
\wh{\R}^{(-,-)}_{0_-,1^*_-}\wh{\MB{R}}^{(-,-)}_{2^*,\ldots,s^*}(\vec{w}_1)
\wh{\BB{T}}^{(-,-)}_{s;0;2^*,\ldots,(Q+1)^*}(\vec{w}_1)\,.
\end{array}
$$
It follows from the definitions of the operators that in order to prove
a statement, it is sufficient to prove the relation
$$
\begin{array}{l}
\tilde{h}(w_s,x)f(w_1,w_s)
\wh{\MB{R}}^{(-,-)}_{1^*_-,s^*_-}(w_s,w_1)\ldots
\wh{\MB{R}}^{(-,-)}_{(s-1)^*_-,s^*_-}(w_s,w_{s-1})
\wh{\MB{R}}^{(-,-)}_{0_-,1^*_-}(w_s,w_1)\ldots
\wh{\R}^{(-,-)}_{0_-,s^*_-}=\\[9pt]
\hskip10mm= \dfrac{\tilde{h}(w_1,x)}{h(w_1,x)}\tilde{h}(w_s,x)
\wh{\MB{R}}^{(-,-)}_{2^*_-,s^*_-}(w_s,w_2)\ldots
\wh{\MB{R}}^{(-,-)}_{(s-1)^*_-,s^*_-}(w_s,w_{s-1})\\[4pt]
\hskip50mm \wh{\MB{R}}^{(-,-)}_{0_-,2^*_-}(w_s,w_2)\ldots
\wh{\R}^{(-,-)}_{0_-,s^*_-}-\\[9pt]
\hskip15mm- \tilde{h}(w_1,x)g(w_s,w_1)
\wh{\R}^{(-,-)}_{1^*_-,s^*_-}\wh{\R}^{(-,-)}_{0_-,1^*_-}
\wh{\MB{R}}^{(-,-)}_{2^*_-,s^*_-}(w_s,w_2)\ldots
\wh{\MB{R}}^{(-,-)}_{(s-1)^*_-,s^*_-}(w_s,w_{s-1})\\[4pt]
\hskip50mm \wh{\MB{R}}^{(-,-)}_{0_-,2^*_-}(w_s,w_2)\ldots
\wh{\R}^{(-,-)}_{0_-,s^*_-}
\end{array}
$$
By direct calculation it is possible to show that the Yang--Baxter
equation
$$
\wh{\MB{R}}^{(-,-)}_{t^*_-,s^*_-}(w_s,w_t)
\wh{\MB{R}}^{(-,-)}_{0_-,t^*_-}(w_s,w_t)
\wh{\R}^{(-,-)}_{0_-,s^*_-}= \wh{\R}^{(-,-)}_{0_-,s^*_-}
\wh{\MB{R}}^{(-,-)}_{0_-,t^*_-}(w_s,w_t)
\wh{\MB{R}}^{(-,-)}_{t^*_-,s^*_-}(w_s,w_t)
$$
holds and by its repeated use we find that for the proof of the
first statement it is sufficient to prove the relation
$$
\begin{array}{l}
\tilde{h}(w_s,x)f(w_1,w_s)
\wh{\MB{R}}^{(-,-)}_{1^*_-,s^*_-}(w_s,w_1)
\wh{\MB{R}}^{(-,-)}_{0_-,1^*_-}(w_s,w_1)
\wh{\R}^{(-,-)}_{0_-,s^*_-}=\\[9pt]
\hskip10mm=
\dfrac{\tilde{h}(w_1,x)}{h(w_1,x)}\tilde{h}(w_s,x)\wh{\R}^{(-,-)}_{0_-,s^*_-}-
\tilde{h}(w_1,x)g(w_s,w_1)\wh{\R}^{(-,-)}_{1^*_-,s^*_-}\wh{\R}^{(-,-)}_{0_-,1^*_-}
\wh{\R}^{(-,-)}_{0_-,s^*_-}\,.
\end{array}
$$
But this relationship is equivalent to the identity
$$
\tilde{h}(w_s,x)f(w_1,w_s)=
\dfrac{\tilde{h}(w_1,x)}{h(w_1,x)}\tilde{h}(w_s,x)-
\tilde{h}(w_1,x)g(w_s,w_1)\,.
$$

\bigskip

To prove the second relationship, we use Lemmas 1 and 3, from which it follows
$$
\begin{array}{l}
T^{-n}_{-n}(x)\Bigl<\MB{b}^{(+)}_{1^*,\ldots,(P+1)^*}(\vec{v}),\MB{e}_{\vec{k}}\Bigr>=
T^{-n}_{-n}(x)\Bigl<\MB{b}^{(+)}_{1^*},\MB{e}_{k_1}\Bigr>
\Bigl<\MB{b}^{(+)}_{2^*,\ldots,(P+1)^*}(\vec{v}_1),\MB{e}_{\vec{k}_1}\Bigr>=\\[6pt]
\hskip10mm= F(x+n-1-\eta;\ol{v})
\Bigl<\MB{b}^{(+)}_{1^*,\ldots,(P+1)^*}(\vec{v}_1),\MB{e}_{\vec{k}}\Bigr>T^{-n}_{-n}(x)+\\[6pt]
\hskip15mm+
\tilde{h}(x,v_1)F(v_1;\ol{v}_1)
\Tr_{0}\biggl(\Bigl<\MB{b}^{(+)}_{2^*,\ldots,(P+1)^*}(\vec{v}_1)\MB{b}^{(-)}_1(x),
\wh{\BB{P}}^{(-,+)}_{1^*_-,1_+}\wh{\BB{T}}^{(+,+)}_{1;0;2,\ldots,P+1}(\vec{v})
\MB{e}_{\vec{k}}\Bigr>\biggr)+\\[6pt]
\hskip15mm+
{\tsum_{v_{\ell}\in\ol{v}_1}}\dfrac{\tilde{h}(x,v_1)}{h(x,v_1)}\tilde{h}(x,v_{\ell})
F(v_{\ell};\ol{v}_{1,\ell})
\Tr_{0}\biggl(\Bigl<\MB{b}^{(+)}_{1^*,\ldots,\wh{\ell}^*,\ldots,(P+1)^*}(\vec{v}_{\ell})
\MB{b}^{(-)}_{\ell}(x),\\[4pt]
\hskip60mm
\wh{\BB{P}}^{(-,+)}_{\ell^*_-,\ell_+}\wh{\MB{R}}^{(+,+)}_{2,\ldots,\ell}(\vec{v}_1)
\wh{\BB{T}}^{(+,+)}_{\ell;0;2,\ldots,P+1}(\vec{v}_1)\MB{e}_{\vec{k}}\Bigr>\biggr)-\\[6pt]
\hskip15mm-
{\tsum_{v_{\ell}\in\ol{v}_1}}\tilde{h}(x,v_1)g(v_1,v_{\ell})F(v_{\ell};\ol{v}_{1,\ell})
\Tr_{0}\biggl(\Bigl<\MB{b}^{(+)}_{\ell^*;2^*,\ldots,(P+1)^*}(v_1;\vec{v}_{1,\ell})\MB{b}^{(-)}_1(x),\\[4pt]
\hskip60mm
\wh{\BB{P}}^{(-,+)}_{1^*_-,1_+}\wh{\MB{R}}^{(+,+)}_{2,\ldots,\ell}(\vec{v}_1)
\wh{\BB{T}}^{(+,+)}_{\ell;0;2,\ldots,P+1}(\vec{v}_1)\wh{\R}^{(+,+)}_{0_+,1_+}
\,\MB{e}_{\vec{k}}\Bigr>\biggr)
\end{array}
$$
According to Lemma 2,
$$
\begin{array}{l}
\Bigl<\MB{b}^{(+)}_{\ell^*;2^*,\ldots,(P+1)^*}(v_1;\vec{v}_{1,\ell})\MB{b}^{(-)}_1(x),
\wh{\BB{P}}^{(-,+)}_{1^*_-,1_+}\wh{\MB{R}}^{(+,+)}_{2,\ldots,\ell}(\vec{v}_1)
\wh{\BB{T}}^{(+,+)}_{\ell;0;2,\ldots,P+1}(\vec{v}_1)\wh{\R}^{(+,+)}_{0_+,1_+}
\,\MB{e}_{\vec{k}}\Bigr>=\\[4pt]
\hskip10mm=
\Bigl<\MB{b}^{(+)}_{1^*,\ldots,\wh{\ell}^*,\ldots,(P+1)^*}(\vec{v}_{\ell})\MB{b}^{(-)}_{\ell}(x),
\wh{\BB{P}}^{(-,+)}_{2\ell^*_-,\ell_+}\wh{\R}^{(+,+)}_{1_+,\ell_+}
\wh{\MB{R}}^{(+,+)}_{2,\ldots,\ell}(\vec{v}_1)
\wh{\BB{T}}^{(+,+)}_{\ell;0;2,\ldots,P+1}(\vec{v}_1)\wh{\R}^{(+,+)}_{0_+,1_+}
\,\MB{e}_{\vec{k}}\Bigr>,
\end{array}
$$
and so
$$
\begin{array}{l}
T^{-n}_{-n}(x)\Bigl<\MB{b}^{(+)}_{1^*,\ldots,(P+1)^*}(\vec{v}),\MB{e}_{\vec{k}}\Bigr>=
F(x+n-1-\eta;\ol{v})
\Bigl<\MB{b}^{(+)}_{1^*,\ldots,(P+1)^*}(\vec{v}_1),\MB{e}_{\vec{k}}\Bigr>T^{-n}_{-n}(x)+\\[6pt]
\hskip15mm+
\tilde{h}(x,v_1)F(v_1;\ol{v}_1)\Tr_{0}\biggl(\Bigl<\MB{b}^{(+)}_{2^*,\ldots,(P+1)^*}(\vec{v}_1)\MB{b}^{(-)}_1(x),
\wh{\BB{P}}^{(-,+)}_{1^*_-,1_+}\wh{\BB{T}}^{(+,+)}_{1;0;2,\ldots,P+1}(\vec{v})
\MB{e}_{\vec{k}}\Bigr>\biggr)+\\[6pt]
\hskip15mm+
{\tsum_{v_{\ell}\in\ol{v}_1}}\dfrac{\tilde{h}(x,v_1)}{h(x,v_1)}\tilde{h}(x,v_{\ell})
F(v_{\ell};\ol{v}_{1,\ell})
\Tr_{0}\biggl(\Bigl<\MB{b}^{(+)}_{1^*,\ldots,\wh{\ell}^*,\ldots,(P+1)^*}(\vec{v}_{\ell})
\MB{b}^{(-)}_{\ell}(x),\\[4pt]
\hskip60mm
\wh{\BB{P}}^{(-,+)}_{\ell^*_-,\ell_+}\wh{\MB{R}}^{(+,+)}_{2,\ldots,\ell}(\vec{v}_1)
\wh{\BB{T}}^{(+,+)}_{\ell;0;2,\ldots,P+1}(\vec{v}_1)\MB{e}_{\vec{k}}\Bigr>\biggr)-\\[6pt]
\hskip15mm-
{\tsum_{v_{\ell}\in\ol{v}_1}}\tilde{h}(x,v_1)g(v_1,v_{\ell})F(v_{\ell};\ol{v}_{1,\ell})
\Tr_{0}\biggl(\Bigl<\MB{b}^{(+)}_{1^*,\ldots,\wh{\ell}^*,\ldots,(P+1)^*}(\vec{v}_{\ell})
\MB{b}^{(-)}_{\ell}(x),\\[4pt]
\hskip60mm
\wh{\BB{P}}^{(-,+)}_{2\ell^*_-,\ell_+}\wh{\R}^{(+,+)}_{1_+,\ell_+}
\wh{\MB{R}}^{(+,+)}_{2,\ldots,\ell}(\vec{v}_1)
\wh{\BB{T}}^{(+,+)}_{\ell;0;2,\ldots,P+1}(\vec{v}_1)\wh{\R}^{(+,+)}_{0_+,1_+}
\,\MB{e}_{\vec{k}}\Bigr>\biggr)
\end{array}
$$
Therefore, it is sufficient to show that the relation
$$
\begin{array}{l}
\tilde{h}(x,v_{\ell})f(v_{\ell},v_1)
\wh{\MB{R}}^{(+,+)}_{1,\ldots,\ell}(\vec{v})
\wh{\BB{T}}^{(+,+)}_{\ell;0;1,\ldots,P+1}(\vec{v})=\\[9pt]
\hskip20mm= \dfrac{\tilde{h}(x,v_1)}{h(x,v_1)}\tilde{h}(x,v_{\ell})
\wh{\MB{R}}^{(+,+)}_{2,\ldots,\ell}(\vec{v}_1)
\wh{\BB{T}}^{(+,+)}_{\ell;0;2,\ldots,P+1}(\vec{v}_1)-\\[9pt]
\hskip40mm- \tilde{h}(x,v_1)g(v_1,v_{\ell})
\wh{\R}^{(+,+)}_{1_+,\ell_+}\wh{\MB{R}}^{(+,+)}_{2,\ldots,\ell}(\vec{v}_1)
\wh{\BB{T}}^{(+,+)}_{\ell;0;2,\ldots,P+1}(\vec{v}_1)\wh{\R}^{(+,+)}_{0_+,1_+}
\end{array}
$$
is valid for any $\ell=2,\,\ldots,\,P+1$.

When we use the definitions
$\wh{\MB{R}}^{(+,+)}_{1,\ldots,\ell}(\vec{v})$,
$\wh{\MB{R}}^{(+,+)}_{2,\ldots,\ell}(\vec{v})$,
$\wh{\BB{T}}^{(+,+)}_{\ell;0;1,\ldots,P+1}(\vec{v})$ and
$\wh{\BB{T}}^{(+,+)}_{\ell;0;2,\ldots,P+1}(\vec{v}_1)$, we find that
to prove the second statement, it is enough to show the equality
$$
\begin{array}{l}
\tilde{h}(x,v_{\ell})f(v_{\ell},v_1)
\wh{\MB{R}}^{(+,+)}_{1,\ell}(v_1,v_\ell)\ldots
\wh{\MB{R}}^{(+,+)}_{\ell-1,\ell}(v_{\ell-1},v_\ell)
\wh{\R}^{(+,+)}_{0_+,\ell_+}\\[4pt]
\hskip60mm
\wh{\MB{R}}^{(+,+)}_{0_+,(\ell-1)_+}(v_{\ell},v_{\ell-1})\ldots
\wh{\MB{R}}^{(+,+)}_{0_+,1_+}(v_{\ell},v_1)=\\[6pt]
\hskip10mm= \dfrac{\tilde{h}(x,v_1)}{h(x,v_1)}\tilde{h}(x,v_{\ell})
\wh{\MB{R}}^{(+,+)}_{2,\ell}(v_2,v_\ell)\ldots
\wh{\MB{R}}^{(+,+)}_{\ell-1,\ell}(v_{\ell-1},v_\ell)
\wh{\R}^{(+,+)}_{0_+,\ell_+}\\[4pt]
\hskip60mm
\wh{\MB{R}}^{(+,+)}_{0_+,(\ell-1)_+}(v_{\ell},v_{\ell-1})\ldots
\wh{\MB{R}}^{(+,+)}_{0_+,2_+}(v_{\ell},v_2)-\\[6pt]
\hskip20mm- \tilde{h}(x,v_1)g(v_1,v_{\ell})
\wh{\R}^{(+,+)}_{1_+,\ell_+}
\wh{\MB{R}}^{(+,+)}_{2,\ell}(v_2,v_\ell)\ldots
\wh{\MB{R}}^{(+,+)}_{\ell-1,\ell}(v_{\ell-1},v_\ell)
\wh{\R}^{(+,+)}_{0_+,\ell_+}\\[4pt]
\hskip60mm
\wh{\MB{R}}^{(+,+)}_{0_+,(\ell-1)_+}(v_{\ell},v_{\ell-1})\ldots
\wh{\MB{R}}^{(+,+)}_{0_+,2_+}(v_{\ell},v_2)
\wh{\R}^{(+,+)}_{0_+,1_+}\,.
\end{array}
$$
By repeatedly using the Yang--Baxter equation
$$
\wh{\MB{R}}^{(+,+)}_{k_+,\ell_+}(v_k,v_\ell)
\wh{\R}^{(+,+)}_{0_+,\ell_+}
\wh{\MB{R}}^{(+,+)}_{0_+,k_+}(v_{\ell},v_k)=
\wh{\MB{R}}^{(+,+)}_{0_+,k_+}(v_{\ell},v_k)
\wh{\R}^{(+,+)}_{0_+,\ell_+}
\wh{\MB{R}}^{(+,+)}_{k_+,\ell_+}(v_k,v_\ell)\,,
$$
which can be verified by direct calculation, we find that to prove
the statement it is enough to prove the relation
$$
\begin{array}{l}
\tilde{h}(x,v_{\ell})f(v_{\ell},v_1)
\wh{\MB{R}}^{(+,+)}_{1,\ell}(v_1,v_\ell)
\wh{\R}^{(+,+)}_{0_+,\ell_+}
\wh{\MB{R}}^{(+,+)}_{0_+,1_+}(v_{\ell},v_1)=\\[4pt]
\hskip20mm=
\dfrac{\tilde{h}(x,v_1)}{h(x,v_1)}\tilde{h}(x,v_{\ell})\wh{\R}^{(+,+)}_{0_+,\ell_+}-
\tilde{h}(x,v_1)g(v_1,v_{\ell})\wh{\R}^{(+,+)}_{1_+,\ell_+}\wh{\R}^{(+,+)}_{0_+,\ell_+}
\wh{\R}^{(+,+)}_{0_+,1_+}
\end{array}
$$
which is equivalent to the identity
$$
\tilde{h}(x,v_{\ell})f(v_{\ell},v_1)=
\dfrac{\tilde{h}(x,v_1)}{h(x,v_1)}\tilde{h}(x,v_{\ell})-
\tilde{h}(x,v_1)g(v_1,v_{\ell})\,.
$$

\bigskip

Assuming that the third statement holds for $Q$, we get by Lemmas 1
and 3
$$
\begin{array}{l}
\tilde{\MB{T}}^{(+)}_0(x)\Bigl<\MB{b}^{(-)}_{1,\ldots,Q+1}(\vec{w}),\MB{f}^{-\vec{r}}\Bigr>=
\Bigl<\MB{b}^{(-)}_{1,\ldots,Q+1}(\vec{w}),
\wh{\MB{T}}^{(+,-)}_{0;1,\ldots,Q+1}(x;\vec{w})\MB{f}^{-\vec{r}}\Bigr>-\\[6pt]
\hskip15mm-
\tilde{h}(w_1,x)F(w_1;\ol{w}_1)
\Bigl<\MB{b}^{(+)}_{1^*}(x)\MB{b}^{(-)}_{2,\ldots,Q+1}(\vec{w}_1),
\wh{\BB{P}}^{(+,-)}_{1_+,1^*_-}\wh{\R}^{(+,-)}_{0_+,1^*_-}\MB{f}^{-\vec{r}}\Bigr>T^{-n}_{-n}(w_1)-\\[6pt]
\hskip15mm-
{\tsum_{w_s\in\ol{w}_1}}\tilde{h}(w_s,x)F(w_s;\ol{w}_{1,s})
\Bigl<\MB{b}^{(+)}_{s^*}(x)\MB{b}^{(-)}_{1,\ldots,\wh{s},\ldots,Q+1}(\vec{w}_s),\\[4pt]
\hskip40mm
\wh{\BB{P}}^{(+,-)}_{s_+,s^*_-}\wh{\MB{R}}^{(+,-)}_{0_+,1^*_-}(x,w_1)\wh{\R}^{(+,-)}_{0_+,s^*_-}
\wh{\MB{R}}^{(-,-)}_{2^*,\ldots,s^*}(\vec{w}_1)
\MB{f}^{-\vec{r}}\Bigr>T^{-n}_{-n}(w_s)+\\[6pt]
\hskip15mm+
{\tsum_{w_s\in\ol{w}_1}}\tilde{h}(w_1,x)g(w_1,w_s)F(w_s;\ol{w}_{1,s})
\Bigl<\MB{b}^{(+)}_{1^*}(x)\MB{b}^{(-)}_{s;2,\ldots,Q+1}(w_1;\vec{w}_{1,s}),\\[4pt]
\hskip40mm
\wh{\BB{P}}^{(+,-)}_{1_+,1^*_-}\wh{\R}^{(+,-)}_{0_+,1^*_-}
\wh{\MB{R}}^{(-,-)}_{2^*,\ldots,s^*}(\vec{w}_1)\MB{f}^{-\vec{r}}\Bigr>T^{-n}_{-n}(w_s)
\end{array}
$$
According to Lemma 2,
$$
\begin{array}{l}
\Bigl<\MB{b}^{(+)}_{1^*}(x)\MB{b}^{(-)}_{s;2,\ldots,Q+1}(w_1;\vec{w}_{1,s}),
\wh{\BB{P}}^{(+,-)}_{1_+,1^*_-}\wh{\R}^{(+,-)}_{0_+,1^*_-}
\wh{\MB{R}}^{(-,-)}_{2^*,\ldots,s^*}(\vec{w}_1)\MB{f}^{-\vec{r}}\Bigr>=\\[6pt]
\hskip10mm=
\Bigl<\MB{b}^{(+)}_{s^*}(x)\MB{b}^{(-)}_{1,\ldots,\wh{s},\ldots,Q+1}(w_1;\vec{w}_s),
\wh{\BB{P}}^{(+,-)}_{s_+,s^*_-}\wh{\R}^{(-,-)}_{1^*_-,s^*_-}
\wh{\R}^{(+,-)}_{0_+,1^*_-}
\wh{\MB{R}}^{(-,-)}_{2^*,\ldots,s^*}(\vec{w}_1)\MB{f}^{-\vec{r}}\Bigr>
\end{array}
$$
and so
$$
\begin{array}{l}
\tilde{\MB{T}}^{(+)}_0(x)\Bigl<\MB{b}^{(-)}_{1,\ldots,Q+1}(\vec{w}),\MB{f}^{-\vec{r}}\Bigr>=
\Bigl<\MB{b}^{(-)}_{1,\ldots,Q+1}(\vec{w}),
\wh{\MB{T}}^{(+,-)}_{0;1,\ldots,Q+1}(x;\vec{w})\MB{f}^{-\vec{r}}\Bigr>-\\[6pt]
\hskip15mm-
\tilde{h}(w_1,x)F(w_1;\ol{w}_1)\Bigl<\MB{b}^{(+)}_{1^*}(x)\MB{b}^{(-)}_{2,\ldots,Q+1}(\vec{w}_1),
\wh{\BB{P}}^{(+,-)}_{1_+,1^*_-}\wh{\R}^{(+,-)}_{0_+,1^*_-}\MB{f}^{-\vec{r}}\Bigr>T^{-n}_{-n}(w_1)-\\[6pt]
\hskip15mm-
{\tsum_{w_s\in\ol{w}_1}}\tilde{h}(w_s,x)F(w_s;\ol{w}_{1,s})
\Bigl<\MB{b}^{(+)}_{s^*}(x)\MB{b}^{(-)}_{1,\ldots,\wh{s},\ldots,Q+1}(\vec{w}_s),\\[4pt]
\hskip40mm
\wh{\BB{P}}^{(+,-)}_{s_+,s^*_-}\wh{\MB{R}}^{(+,-)}_{0_+,1^*_-}(x,w_1)\wh{\R}^{(+,-)}_{0_+,s^*_-}
\wh{\MB{R}}^{(-,-)}_{2^*,\ldots,s^*}(\vec{w}_1)
\MB{f}^{-\vec{r}}\Bigr>T^{-n}_{-n}(w_s)+\\[6pt]
\hskip15mm+
{\tsum_{w_s\in\ol{w}_1}}\tilde{h}(w_1,x)g(w_1,w_s)F(w_s;\ol{w}_{1,s})
\Bigl<\MB{b}^{(+)}_{s^*}(x)\MB{b}^{(-)}_{1,\ldots,\wh{s},\ldots,Q+1}(w_1;\vec{w}_s),\\[4pt]
\hskip40mm
\wh{\BB{P}}^{(+,-)}_{s_+,s^*_-}\bigl(\tilde{\R}^{(-,-)}_{1^*_-,s^*_-}\bigr)^*
\wh{\R}^{(+,-)}_{0_+,1^*_-}
\wh{\MB{R}}^{(-,-)}_{2^*,\ldots,s^*}(\vec{w}_1)\MB{f}^{-\vec{r}}\Bigr>T^{-n}_{-n}(w_s)
\end{array}
$$
Therefore, it is enough to show that for any $s=2,\ldots,Q+1$ we have
$$
\begin{array}{l}
\tilde{h}(w_s,x)f(w_s,w_1)\wh{\R}^{(+,-)}_{0_+,s^*_-}
\wh{\MB{R}}^{(-,-)}_{1^*_-,s^*_-}(w_s,w_1)=\\[6pt]
\hskip20mm=
\tilde{h}(w_s,x)\wh{\MB{R}}^{(+,-)}_{0_+,1^*_-}(x,w_1)\wh{\R}^{(+,-)}_{0_+,s^*_-}-
\tilde{h}(w_1,x)g(w_1,w_s)\wh{\R}^{(-,-)}_{1^*_-,s^*_-}
\wh{\R}^{(+,-)}_{0_+,1^*_-}
\end{array}
$$
If we use the definitions of these mappings, we find that these
relations are equivalent to identity
$$
\tilde{h}(w_s,x)g(w_s,w_1)+\tilde{h}(w_s,x)\tilde{h}(w_1,x)+\tilde{h}(w_1,x)g(w_1,w_s)=0\,.
$$

\bigskip

To prove the fourth statement, we use the relation
$$
%
$$
which follows from Lemma 1 and the inductive assumption. According
to Lemma 2, we have
$$
\begin{array}{l}
\Bigl<\MB{b}^{(+)}_{\ell^*;2^*,\ldots,(P+1)^*}(v_1;\vec{v}_{1,\ell})\MB{b}^{(-)}_1(x),
\wh{\BB{P}}^{(-,+)}_{1^*_-,1_+}\wh{\R}^{(-,+)}_{0_-,1_+}
\wh{\MB{R}}^{(+,+)}_{2,\ldots,\ell}(\vec{v}_1)\MB{e}_{\vec{k}}\Bigr>=\\[4pt]
\hskip20mm=
\Bigl<\MB{b}^{(+)}_{1^*,\ldots,\wh{\ell}^*,\ldots,(P+1)^*}(\vec{v}_\ell)\MB{b}^{(-)}_1(x),
\wh{\BB{P}}^{(-,+)}_{\ell^*_-,\ell_+}\wh{\R}^{(+,+)}_{1_+,\ell_+}
\wh{\R}^{(-,+)}_{0_-,1_+}
\wh{\MB{R}}^{(+,+)}_{2,\ldots,\ell}(\vec{v}_1)\MB{e}_{\vec{k}}\Bigr>
\end{array}%
$$
Therefore, it is enough to show that equality
$$
\begin{array}{l}
\tilde{h}(x,v_{\ell})f(v_1,v_{\ell}) \wh{\R}^{(-,+)}_{0_-,\ell_+}
\wh{\MB{R}}^{(+,+)}_{1,\ell}(v_1,v_\ell)=\\[4pt]
\hskip10mm= \tilde{h}(x,v_{\ell}) \wh{\R}^{(-,+)}_{0_-,\ell_+}
\wh{\MB{R}}^{(-,+)}_{0_-,1_+}(x,v_1)- \tilde{h}(x,v_1)g(v_\ell,v_1)
\wh{\R}^{(+,+)}_{1_+,\ell_+} \wh{\R}^{(-,+)}_{0_-,1_+}\,.
\end{array}
$$
holds for any $\ell=2,\,\ldots,\,P+1$. And if we use the definitions
of the involved operators, we find that this equality is
equivalent to the relation
$$
\tilde{h}(x,v_{\ell})g(v_1,w_\ell)+\tilde{h}(x,v_{\ell})\tilde{h}(x,v_1)+\tilde{h}(x,v_1)g(v_\ell,v_1)=0\,,
$$
which can be easily verified. \qed

\bigskip

\underbar{\textbf{Lemma 5.}} For any $\vec{v}$, $\vec{w}$, $\vec{k}$
and $\vec{r}$ the relations
$$
%
$$
hold.

\medskip

\textsc{Proof:} In proving this Lemma, we will often use the equation
$$
%
$$
which follows from the relation
$\MB{b}^{(+)}_{1^*}(x)\MB{b}^{(-)}_2(y)=\MB{b}^{(-)}_2(y)\MB{b}^{(+)}_{1^*}(x)$.
We calculate action of the operators $T^{\pm n}_{\pm n}(x)$ and
$\tilde{\MB{T}}^{(\pm)}_0(x)$ on the element
$\bigl<\MB{b}^{(+)}_{1^*,\ldots,P^*}(\vec{v})\MB{b}^{(-)}_{1,\ldots,Q}(\vec{w}),
\MB{e}_{\vec{k}}\otimes\MB{f}^{-\vec{r}}\bigr>$ in both orders and
then we get assertion of Lemma 5 by comparing these expressions.

From Lemmas 3 and 4 we get
$$
%
$$
In the first term the vectors $\vec{v}$ and $\vec{w}$ do not change,
in the second the components $v_\ell$ and $x$ are interchanged
and the third contains expressions in which $w_s$ and $x$ are interchanged.

On the other hand, we also have
$$
%
$$
In this expression the vectors $\vec{v}$ and $\vec{w}$ do not change
in the first term, in the second $x$ is changed by $v_\ell$ and
in the third $x$ and $w_s$ are interchanged.
If we compare these two expressions, we get the first statement of
Lemma 5.

We get the second relation when we compare the equalities
$$
%
$$

The third equality is obtained by comparing the equalities
$$
%
$$
and the fourth relation is the result of equalities
$$
%
$$
\qed

\end{document}